\documentclass[aps,prx,twocolumn,english,superscriptaddress,floatfix,longbibliography,nofootinbib,10pt]{revtex4-2}

\usepackage{dsfont}

\usepackage{amsmath}
\usepackage{mathtools}
\usepackage{amssymb}
\usepackage{xcolor}
\usepackage{nicefrac}
\usepackage{bbold}
\usepackage{braket}
\usepackage{hhline}
\usepackage{lipsum}
\usepackage{booktabs}
\usepackage{multirow}
\usepackage{hhline}
\usepackage{rotating}
\usepackage{adjustbox}

\usepackage{physics}

\usepackage{amsfonts}
\usepackage{stmaryrd}

\usepackage{xparse}
\usepackage{graphicx}
\definecolor{mygreen}{rgb}{0.328125,0.6796875,0.1953125}
\definecolor{myblue}{rgb}{0.12156862745098039, 0.4666666666666667, 0.7058823529411765}
\definecolor{babyblue}{rgb}{0.42, 0.84, 1}
\usepackage{hyperref}
\hypersetup{
     colorlinks=true,
     linkcolor=myblue,
     filecolor=magenta,      
     urlcolor=myblue,
     citecolor=myblue
}

\usepackage{enumitem}
\usepackage{tikz}

\usepackage{ragged2e}

\newcommand{\cz}[2]{\ifmmode \mathrm{C}_{#1}\mathrm{Z}_{#2} \else $\mathrm{C}_{#1}\mathrm{Z}_{#2}$\fi}

\NewDocumentCommand{\cnot}{o o}{
  \ensuremath{
    C\IfValueT{#1}{_{#1}}X\IfValueT{#2}{_{#2}}
  }
}

\NewDocumentCommand{\apcp}{o}{
  \ensuremath{
    \mathrm{a}PCP\IfValueT{#1}{_{#1}}
  }
}

\NewDocumentCommand{\lcnot}{o o}{
  \ensuremath{
    \overline{C\IfValueT{#1}{_{#1}}X\IfValueT{#2}{_{#2}}}
  }
}

\NewDocumentCommand{\lcx}{o o}{
  \ensuremath{
    \overline{C\IfValueT{#1}{_{#1}}X\IfValueT{#2}{_{#2}}}
  }
}

\NewDocumentCommand{\ls}{o}{
  \ensuremath{
    \overline{S}\IfValueT{#1}{_{#1}}
    }
}

\NewDocumentCommand{\lh}{o}{
  \ensuremath{
    \overline{H}\IfValueT{#1}{_{#1}}
  }
}

\NewDocumentCommand{\lec}{o}{
  \ensuremath{
    \overline{EC}\IfValueT{#1}{_{#1}}
  }
}

\NewDocumentCommand{\lx}{o}{
  \ensuremath{
    \overline{X}\IfValueT{#1}{_{#1}}
  }
}
\NewDocumentCommand{\ly}{o}{
  \ensuremath{
    \overline{Y}\IfValueT{#1}{_{#1}}
  }
}

\NewDocumentCommand{\lz}{o}{%
  \ensuremath{
    \overline{Z}\IfValueT{#1}{_{#1}}%
  }
}

\newcommand\brpm{\mathbin{\vcenter{\hbox{\oalign{$\scriptstyle({+})$\cr
					\noalign{\kern-.3ex}
					\hfil$\scriptscriptstyle-$\hfil\cr}}}}}

\let\originalleft\left
\let\originalright\right
\renewcommand{\left}{\mathopen{}\mathclose\bgroup\originalleft}
\renewcommand{\right}{\aftergroup\egroup\originalright}

\DeclareMathOperator{\supp}{supp}

\begin{document}

\title{Addressable fault-tolerant universal quantum gate operations for high-rate \hspace{1cm} lift-connected surface codes}

\author{Josias Old}
    \email{j.old@fz-juelich.de}
    \affiliation{Institute for Quantum Information, RWTH Aachen University, Aachen, Germany}
    \affiliation{Institute for Theoretical Nanoelectronics (PGI-2), Forschungszentrum J\"{u}lich, J\"{u}lich, Germany}
    \affiliation{\href{https://www.neqxt.org/}{\color{black} neQxt GmbH, Cologne, Germany}}

\author{Juval Bechar}
    \affiliation{Institute for Quantum Information, RWTH Aachen University, Aachen, Germany}
    \affiliation{Institute for Theoretical Nanoelectronics (PGI-2), Forschungszentrum J\"{u}lich, J\"{u}lich, Germany}

\author{Markus Müller}
    \affiliation{Institute for Quantum Information, RWTH Aachen University, Aachen, Germany}
    \affiliation{Institute for Theoretical Nanoelectronics (PGI-2), Forschungszentrum J\"{u}lich, J\"{u}lich, Germany}

\author{Sascha Heu\ss en}
    \email{s.heussen@neqxt.org}
    \affiliation{\href{https://www.neqxt.org/}{\color{black} neQxt GmbH, Cologne, Germany}}

\begin{abstract}

Quantum low-density parity check (qLDPC) codes are among the leading candidates to realize error-corrected quantum memories with low qubit overhead. Potentially high encoding rates and large distance relative to their block size make them appealing for practical suppression of noise in near-term quantum computers. In addition to increased qubit-connectivity requirements compared to more conventional topological quantum error correcting codes, qLDPC codes remain notoriously hard to compute with. In this work, we introduce a construction to implement all Clifford quantum gate operations on the recently introduced lift-connected surface (LCS) codes\,\cite{old2024lift}. These codes can be implemented in a 3D-local architecture and achieve asymptotic scaling $[[n, \mathcal{O}(n^{1/3}), \mathcal{O}(n^{1/3})]]$. In particular, LCS codes realize favorable instances with small numbers of qubits: For the $[[15,3,3]]$ LCS code, we provide deterministic fault-tolerant (FT) circuits of the logical gate set $\{\lh[i], \ls[i], \lcnot[i][j]\}_{i,j \in (0,1,2)}$  based on flag qubits. By adding a procedure for FT magic state preparation, we show quantitatively how to realize an FT universal gate set in $d=3$ LCS codes. Numerical simulations indicate that our gate constructions can attain pseudothresholds in the range $p_{\mathrm{th}} \approx 4.8\cdot 10^{-3}-1.2\cdot 10^{-2}$ for circuit-level noise. The schemes use a moderate number of qubits and are therefore feasible for near-term experiments, facilitating progress for fault-tolerant error corrected logic in high-rate qLPDC codes.

\end{abstract}

\maketitle

\section{Introduction}

Quantum error correcting (QEC) codes provide a means to practically realize fault-tolerant (FT) quantum computation (QC) by overcoming the limitations that are placed by noisy qubits and imperfect physical quantum gate operations performed on them. 
The most prominent QEC codes, like topological surface or color codes, typically have vanishing \emph{rate}, meaning that the relative amount of logical information they encode approaches zero upon increased protection against noise\,\cite{dennis2002topological, bombin2006topological}. High-rate quantum low-density parity check (qLDPC) codes emerged as promising candidates for improved memory capabilities while also reducing the physical qubit overhead compared to, for instance, the paradigmatic surface code\,\cite{tillich2013quantum, gottesman2014fault, breuckmann2021quantum, xu2024constant, Bravyi_2024, Pecorari_2025}. Since the actual goal is FTQC, there is a need for QEC codes that, while achieving high rate and small memory overhead, also allow one to perform FT logical quantum gate operations.

The simplest way to construct a FT logical gate is a \emph{transversal} implementation, where gates only act on single (or bounded number of) disjoint data qubits within a block\,\cite{gottesman1997thesis}. This ensures that not too many faults occur on the data qubits and prevents faults from spreading uncontrollably.
There are, however, only very few error-correcting codes that support transversal Clifford gates. Most notably, the full Clifford group is transversal on 2D topological color codes\,\cite{bombin2006topological}. 
If an error correcting code has a certain structure or symmetries, unitary implementations of gates that are not strictly transversal can still be fault-tolerant. For instance, the logical Hadamard gate can be realized transversally up to qubit permutation in the presence of a so-called $ZX$-duality \,\cite{quintavalle2023partitioning, breuckmann2024fold,eberhardt2024logicaloperatorsfoldtransversalgates,sayginel2025fault}. 
In high-rate codes such gates are often global; they act on all logical qubits of a block and are not able to address individual logical qubits\,\cite{liang2025self}.

A well-established method of achieving addressability of logical qubits is through teleportation-based gates using joint logical measurements\,\cite{brun2015teleportation}. 
The main challenge of this approach is to realize FT logical measurements and has been tackled using lattice-surgery approaches\,\cite{horsman2012surface}. 
Generalizations of the original surface code lattice-surgery employ special but rather involved LDPC ancillary states that enable a decomposition of logical operators into low-weight measurements\,\cite{cohen2022low, williamson2024low,cowtan2024ssip,ide2025fault,cross2024improvedqldpcsurgerylogical,he2025extractorsqldpcarchitecturesefficient,malcolm2025computingefficientlyqldpccodes,baspin2025fast, xu2025batched}.

Between the previous two approaches lies a third strategy to realize FT logical gates. It is based on unitary logical gates implemented with non-FT circuits. Through additional means, the circuits are then rendered FT subsequently. Examples include flag measurements\,\cite{chao2018quantum}, concatenation or intermediate stabilizer measurements in the pieceable fault-tolerance framework\,\cite{yoder2016universal}. This is the approach we follow in our work.
In order to achieve universality, it is important to note that no error-correcting code can implement a \emph{universal} gate set using only transversal operations\,\cite{eastin2009restrictions}.
Implementing a FT universal gate can be challenging. A widely used technique relies on magic state injection. 
There, good quality magic states are used as a non-stabilizer resource to feed into a gate teleportation circuit that itself again only contains Clifford gates but facilitates the logical non-Clifford gate\,\cite{gottesman1999demonstrating,bravyi2005universal}.

Recent experimental implementations of QEC have been focusing mostly on demonstrating increased qubit lifetimes via quantum memory experiments in small code instances\,\cite{egan2021fault, ryan2021realization,postler2022demonstration,krinner2022realizing, zhao2022realization, ai2024quantum, postler2024demonstration}. 
Primitives for a universal set of quantum gate operations have been demonstrated with topological color codes, which are especially favorable in this regard\,\cite{ryananderson2022implementingfaulttolerantentanglinggates, postler2022demonstration,lacroix2025scaling,sales2025experimental,daguerre2025experimental,dasu2025breaking}. 

\begin{figure}
    \centering
    \includegraphics[width=\linewidth]{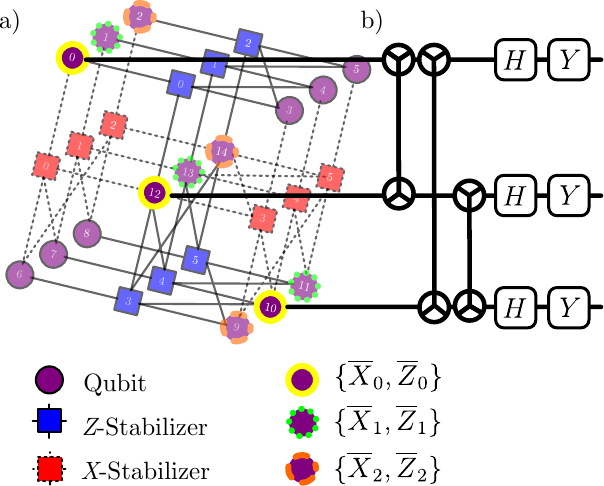}
    \caption{a) Tanner graph of $\llbracket 15,3,3 \rrbracket$ Lift-connected surface code (LCS) and round-robin logical $\lh$ gate. Circles represent data qubits, blue and solid squares the $Z$-stabilizers and red and dashed squares the $X$-stabilizers. This LCS code corresponds to three interconnected copies of $d=2$ surface codes. We indicate the support of the three logical $\overline{X}$- and logical $\overline{Z}$- operators by a yellow solid, green dotted and orange dashed border, respectively. b) For one logical operator we also show the targeted round-robin logical $\lh$ gate. It consists of all-to-all $YCY$ and transversal physical $H$ and $Y$ gates on qubits in the support of the logical operator. In the main text, we explain how to render this logical gate circuit fault-tolerant.}
    \label{fig:lcs_1_3_hgate}
\end{figure}

The main goal of this work is to construct circuits for \emph{addressable} and \emph{fault-tolerant} logical quantum gates in \emph{small} instances of \emph{high-rate} qLDPC codes, which can be realized in near-term experiments.
A particular class of qLDPC codes based on a quasi-cyclic lifted product construction are lift-connected surface (LCS) codes \,\cite{old2024lift}. They consist of sparsely-interconnected copies of 2D surface code patches and can be implemented in a 3D-local fashion.  
The $\llbracket 15,3,3 \rrbracket$ LCS code encodes $3$ logical qubits into $15$ physical qubits and can correct any single Pauli error. Figure \ref{fig:lcs_1_3_hgate}\,a) depicts a visualization of the code. It is made up of $3$ interconnected copies of distance $d=2$ error-detecting surface codes. 
Compared to other qLDPC code constructions, in particular asymptotically good\,\cite{breuckmann2021balanced,panteleev2022asymptotically} or bivariate bicycle codes\,\cite{Bravyi_2024}, LCS codes fall short with respect to asymptotic parameters: $\llbracket n,k,d \rrbracket$ of LCS codes asymptotically scale like disjoint copies of surface codes. 
They offer, however, a well defined constructive family of QEC codes with attractive small to intermediate qubit number members. In a memory setting, these have been shown to achieve similar logical error rates as surface codes while requiring up to four times fewer physical qubits\,\cite{old2024lift}.

In this work, we systematically construct all quantum gate operations required to span the logical Clifford group in LCS codes. We show such a gate operation, a logical Hadamard gate, acting on one of the three logical qubits, in Fig.\,\ref{fig:lcs_1_3_hgate}\,b).
For the $\llbracket 15,3,3 \rrbracket$ code, the logical gate set $\{\lh, \ls, \lcnot\}$ is rendered FT via adding a small number of \emph{flag} qubits to our construction. Our flag-FT Clifford gates are fully deterministic and no postselection on mid-circuit measurement results is required. By also explicitly providing an FT magic state that can facilitate a logical $\pi/4$-rotation, we complete the universal gate set \emph{Clifford+T}. 

This work is structured as follows. In Sec.\,\ref{sec:notionsft}, we introduce the necessary background on error correcting codes and fault tolerance. Sec.~\ref{sec:roundrobin} discusses the synthesis of circuits implementing logical Clifford gates targeting arbitrary (pairs of) logical qubits within LCS codes. All gates that we discover this way respect a round-robin structure. Subsequently in Sec.~\ref{sec:flags}, we lever insights from Ref.\,\cite{chao2018fault} to add flag qubits to our round-robin gates that effectuate our FT Clifford gate set. The performance of our FT gate set, including FT magic state preparation, under circuit-level noise, is scrutinized in numerical simulations in Sec.~\ref{sec:numerics}. We provide conclusions and an outlook on future work in Sec.~\ref{sec:conclusion}.

\section{Error Corrected Quantum Circuits and Fault Tolerance}\label{sec:notionsft}
We consider circuits that can correct errors by implementing a stabilizer code $\mathcal{C}$\,\cite{gottesman1997thesis, nielsen2010quantum}. A quantum error correcting code encoding $k$ logical qubits is defined by a generating set of stabilizer operators $\mathcal{S} = \{S_i\}_{i=0}^{n-k-1} \subseteq \mathcal{P}^{\otimes n} \setminus \{-I\}$, an Abelian subgroup of the $n$-qubit Pauli group. 
The code space is defined as the set of $+1$ eigenstates of all stabilizer generators. 
Encoded Pauli operators are generated by the normalizer of $\mathcal{S}$ in $\mathcal{P}^{\otimes n}$, $\overline{\mathcal{P}} = \mathcal{N}_{\mathcal{P}^{\otimes n}}(\mathcal{S}) \setminus \mathcal{S}$. These operators preserve the stabilizer group, but they are not stabilizers. Therefore, they act nontrivially in the codespace. The minimum weight, i.e.~number of non-identity Pauli terms, of any logical operator is the \emph{distance} $d$ of the code.
A QEC code with distance $d$ can correct any error of weight up to $t = \lfloor \frac{d-1}{2} \rfloor$. 

We adapt a circuit-centric perspective on error correction. Typically, the realization of an error-correcting code induces constraints on the possible measurement outcomes of the corresponding circuit. As an example, measuring a stabilizer operator on the $n$ physical qubits used to encode $k$ logical qubits always yields a $+1$ measurement outcome in the absence of noise. Faults in the execution of the circuit can flip the measurement outcomes to $-1$. This perspective is formalized in the frameworks of spacetime codes\,\cite{delfosse2023spacetime}, detector error models\,\cite{derks2025designing} or Pauli webs and Pauli flows in tensor networks\,\cite{bombin2024unifying, magdalena2025xyz}. 
A matrix $D$ encodes the constraints on the measurement outcomes $\mathbf{m} \in \mathbb{F}_2^{n_{\mathrm{m}}}$ such that in the noise-free execution of the circuit, the \emph{detector outcomes} $\mathbf{d} \in \mathbb{F}_2^{n_{\mathrm{D}}} = D \mathbf{m}$ are trivial, $\mathbf{d} = \mathbf{0}$. The (sets of) deterministic measurements are typically referred to as \emph{detectors}.

We define a circuit-level noise model by assigning Pauli noise channels to each faulty location $\ell \in \mathcal{L}$. We refer to a realization of a non-identity term of a noise channel as a (elementary) fault. A \emph{fault path} $F = (P \subseteq \mathcal{P}^{\otimes n}, L \subseteq \mathcal{L})$ of order $w_F = \abs{F}$ is a collection of elementary faults at $w_F$ locations in the circuit. In a uniform noise model with parameter $p$, any order $w_F$ fault path occurs with probability $\mathcal{O}(p^{w_F})$.  We denote the propagation of the fault path to the error $E$ on the data qubits at the end of the circuit by $E(F) \in \mathcal{P}^{\otimes n}$. We call the \emph{syndrome} of the error its commutation relations with the stabilizer generators of the code, $\mathbf{s}(E) = (\langle E,S_i \rangle)_{i=0}^{n-k-1}$. Here $\langle P,P' \rangle = 0$ if $[P,P'] = 0$ and else $1$.

\begin{figure}
    \centering
    \includegraphics[width=\linewidth]{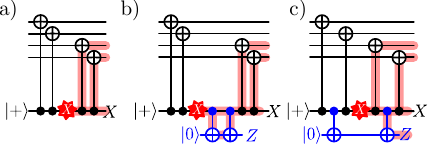}
    \caption{Flagging a $4$-body Pauli-$X$ measurement to catch correlated errors. a) An $X$-fault on the ancilla qubit after the second entangling $CX$ propagates to a weight-2 $XX$ error on the data qubits. b) We insert an identity measurement using an additional flag qubit. This does not change the original measurement, and as such, does not detect the fault. c) Commuting one gate through the second $CX$ can catch the $X$-fault as it now only propagates to the ancilla once and flips the flag-measurement. }
    \label{fig:how2flag}
\end{figure}

A \emph{flag} is an auxiliary qubit that is employed to detect faults that result in correlated errors on data qubits\,\cite{yoder2017the, chao2018quantum}. 
To that end, the flag ancilla implements a projective measurement of the identity, typically using two consecutive controlled-Pauli ($CP$) gates, $(CP)^2 = I$. This flag measurement does not change the (noise-free) action of the circuit, but adds an additional detector to the circuit. 
Commuting one of the gates to another location in the circuit again does not change the action of the circuit. It, however, allows for faults $P' \neq P$ in between the flag gates to propagate to the ancilla and subsequently be detected by the measurement of the flag. 
We show this principle in Fig.\,\ref{fig:how2flag} using a $4$-body $X$-measurement.

\begin{figure}
    \centering
    \includegraphics[width=\linewidth]{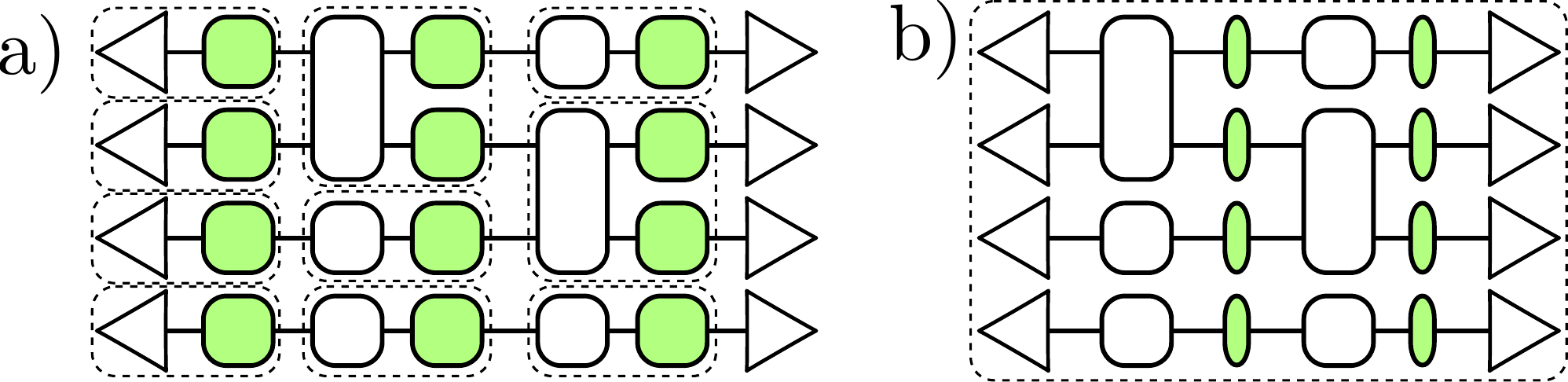}
    \caption{Two notions of fault-tolerant error-correction circuits we consider in this work. We consider a Clifford circuit with encoded state preparation ($\lhd$), logical single- and two qubit gates ($\Box$) and final measurements ($\rhd$). We denote error correction cycles by a green box.
    a) Traditional Gadget Fault Tolerance. Each gadget is independently decoded (dashed rectangle). For overall fault-tolerance, this requires a strictly fault-tolerant implementation of the gates with a fully fault-tolerant round of error correction after each logical gate.
    b) In \emph{Algorithmic Fault Tolerance}, as much information as possible is included in the decoding. For the circuits considered, this can be all detector outcomes of the circuit, including final measurement. This can lead to reduced requirements on overhead of error correction cycles while staying fault tolerant, as indicated by thinner squares. For \emph{transversal algorithmic fault tolerance}, this can be as low as $\Theta(1)$ rounds of syndrome measurements per logical gate\,\cite{zhou2025low}.}
    \label{fig:fig_algFT}
\end{figure}

By detecting more faults, flags can turn circuits \emph{fault-tolerant} (FT). Loosely speaking, an encoded circuit is ($t$-) FT if it is possible to correct for any number of up to $t$ faults. At the same time, faults occurring during the circuit should not propagate uncontrollably in order to prevent accumulation of errors.
In Fig.\,\ref{fig:fig_algFT}, we illustrate two notions of fault-tolerant error correction.

On one side, Fig.\,\ref{fig:fig_algFT}\,a), each physical location is replaced by a fault-tolerant \emph{gadget} that consists of encoded versions of the gates with subsequent fault-tolerant error correction cycles. This has been introduced for concatenated distance-3 codes in Refs.\,\cite{knill1998resilient, aliferis2006quantum}. To achieve fault-tolerance for the full circuit, each circuit element (state preparation, gate or measurement) has to be individually fault-tolerant. 
There are numerous ways to achieve that, e.g.~using encoded ancillae (cat states for Shor-style, logical Pauli states for Steane- and Knill-type EC), $\Theta(d)$ rounds of syndrome measurements or flag-based approaches\,\cite{steane1997active, shor1996fault, divincenzo2007effective, dennis2002topological, chamberland2018flag}. Encoded gates are individually decoded and, e.g., using Monte Carlo simulations, an effective logical error rate can be estimated. For a fault-tolerant implementation using a distance $d$ code, the logical error rate scales as $p_L \propto p^{t+1}$ in the low-$p$ regime.

On the other side, Fig.\,\ref{fig:fig_algFT}\,b), decoding is performed using as much information as possible. For Clifford circuits, this can be all detector outcomes of the circuit. Typically, such simulations show lower logical error rates, mainly for two reasons. The first is that correlations of errors between gadgets can be accounted for. Also, for a final single-qubit measurement data- and measurement-errors are equivalent. In a setting of purely transversal gates, it is also possible to use more compact syndrome measurement circuits ($\Theta(1)$ rounds per logical operation), while keeping fault-tolerance in the sense of suppression of all errors up until order $p^t$. This is referred to as \emph{transversal algorithmic fault-tolerance} in Ref.\,\cite{zhou2025low}.  

In the following, we define fault-tolerant gadgets for $d=3$ similar to Refs.\,\cite{aliferis2006quantum,tansuwannont2022achieving}. We then show fault-tolerant error correction protocols using flags, similar to Ref.\,\cite{chamberland2018flag} and define the corresponding protocols for logical gates. 

The workhorse for these definitions is the property of \emph{distinguishability} of faults, which relaxes traditional proofs considering solely the weights of errors\,\cite{aliferis2006quantum,chamberland2018flag,tansuwannont2022achieving}. 
We say that the set of all fault paths of order up to $t$, $\mathcal{F}^{(t)} = \{F: w_F \leq t\}$, is \emph{distinguishable} if for all pairs of fault paths $F_i, F_j \in \mathcal{F}^{(t)}$ at least one of the following distinguishability criteria is satisfied:

\begin{enumerate}
    \item $\mathbf{d}(F_i) \neq \mathbf{d}(F_j)$:\\ The fault paths lead to different detector outcomes.
    \item $E(F_i) \sim_{\mathcal{S}} E(F_j)$: \\ The fault paths propagate to stabilizer-equivalent errors, i.e.~the errors have the same syndrome but only differ by an element of the stabilizer group.
    \item $\mathbf{s}(E(F_i)) \neq \mathbf{s}(E(F_j))$: \\ Their errors have different syndromes and can therefore be distinguished by a noise-free round of stabilizer measurements.
\end{enumerate}

For a circuit with a distinguishable fault set, a subsequent round of noise-free syndrome measurements allows one to correct all fault paths in $\mathcal{F}^{(t)}$, up to stabilizer-equivalence. We refer to the set of propagated errors of that fault set as $\mathcal{E}^{(t)}$.

\begin{figure}
    \centering
    \includegraphics[width=0.6\linewidth]{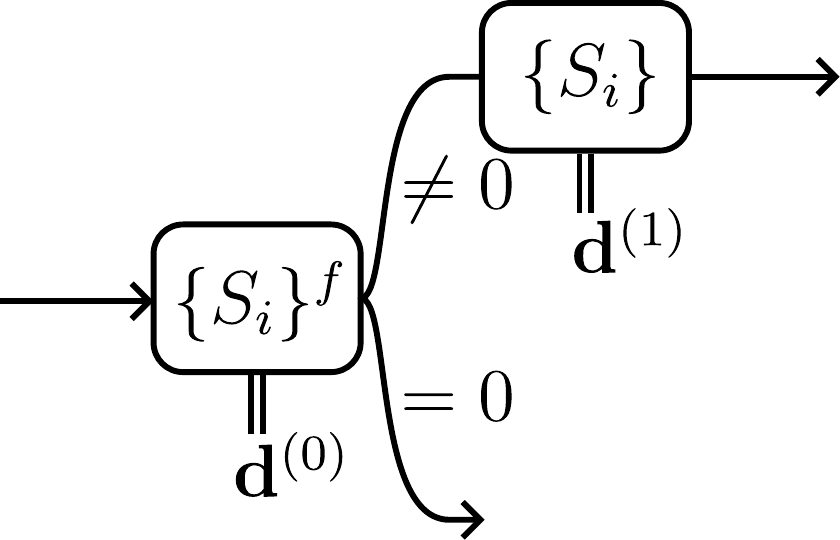}
    \caption{1-Fault-tolerant error-correction protocol. Start by measuring the stabilizer generators with flags $\{S_i\}^f$, resulting in detector outcomes $\mathbf{d}^{(0)}$. If any detector has non-trivial outcome, we know some fault has occurred. A subsequent (non-flagged) stabilizer generator measurement acts as a noise-free round of stabilizer measurements in order $\mathcal{O}(p)$, allowing for unique identification of all $\mathcal{O}(p)$ faults. If all detectors have trivial outcome, the distinguishability of $\{S_i\}^f$ ensures that no or no uncorrectable fault has occurred. In that case, nothing has to be done.}
    \label{fig:ft_ec}
\end{figure}

A fault-tolerant error correction gadget for a code with distance $d = 2t+1 = 3$  has to be constructed such that if no more than $t = 1$ fault occurs, an ideal decoder identifies the correct logical state. 
A simple, but not optimized, protocol that achieves this is shown in Fig.\,\ref{fig:ft_ec}. We denote by $\{S_i\}$ a measurement of all stabilizer generators using single auxiliary qubits. By a superscript $f$, we denote a measurement of all stabilizer generators such that the circuit has a distinguishable fault set $\mathcal{F}^{(1)}$. Typically, a \emph{flagged} stabilizer measurement can achieve that\,\cite{chamberland2018flag}. 

In the $d=3$ case ($t=1$), any non-zero detector outcome of a first flagged syndrome measurement circuit $\{S_i\}^f$ indicates that (at least) one fault has occurred. Then, no more fault will occur in first order $\mathcal{O}(p)$ and we can re-measure the stabilizers without flags to obtain the correct syndrome. 
If all detector outcomes are $0$, the distinguishability of all faults in the flagged syndrome measurement circuit ensures that no uncorrectable fault occurred during the gate. In that case, no subsequent action is necessary.

\begin{figure}
    \centering
    \includegraphics[width=0.8\linewidth]{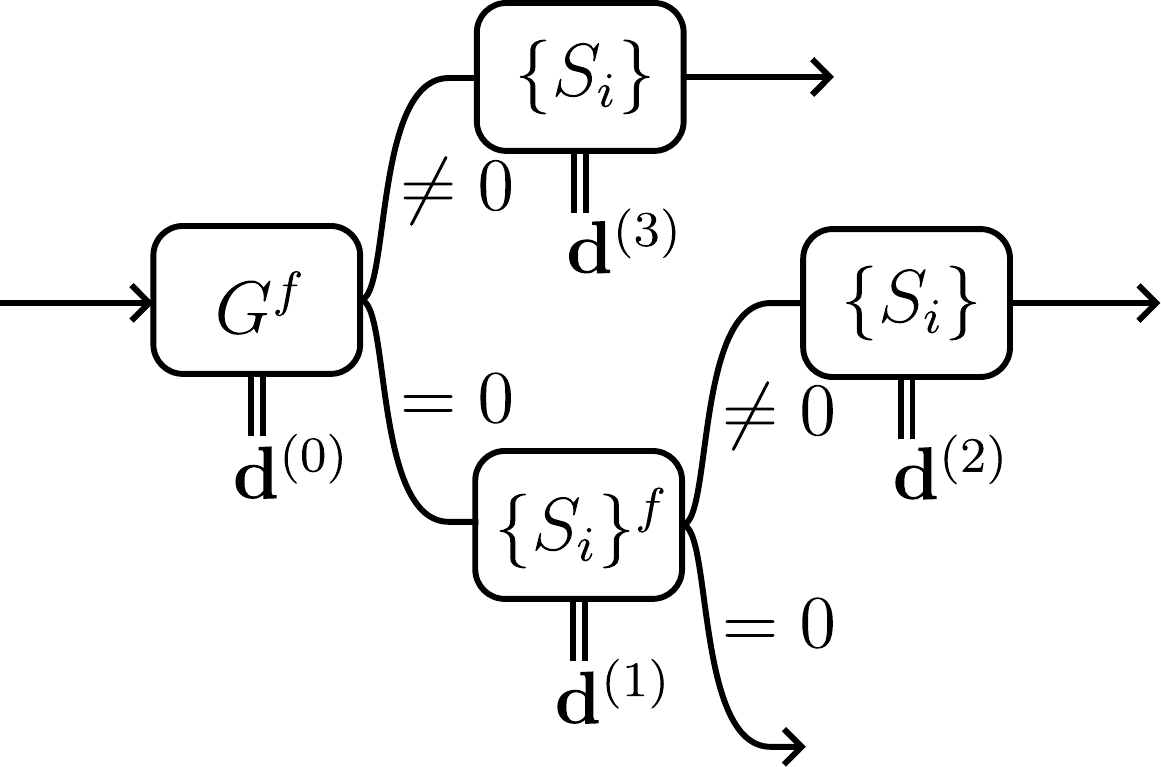}
    \caption{1-Fault-tolerant gate protocol. Start by applying the (flagged) unitary gate $G^f$ that has a distinguishable fault set. If any of the detector outcomes $\mathbf{d}^{(0)}$ is non-trivial, a fault has occurred and a non-flagged round of stabilizer measurement is sufficient to correct all faults in order $\mathcal{O}(p)$. 
    For a trivial detector outcome, we append a round of FT error correction, cf. protocol in Fig.\,\ref{fig:ft_ec}.}
    \label{fig:ft_gate}
\end{figure}

For logical gates, the fault-tolerant gadget acting on $k' \leq k$ encoded qubits $\{i\}$ has to fulfill similar conditions: if no more than $1$ fault occurs, an ideal decoder doesn't induce a logical error.
In contrast to Ref.\,\cite{aliferis2006quantum}, we do not require a bound on spread of error weight within one code-block for fault-tolerance - in our case this is handled by the distinguishability of faults, similar to Ref.\,\cite{tansuwannont2022achieving}. 
In Fig.\,\ref{fig:ft_gate}, we show a simple (but not optimal) protocol to construct a fault-tolerant gate for $d=3$, $t=1$ (\emph{1-fault-tolerant}). Again, if the fault set $\mathcal{F}^{(1)}$ is distinguishable in the flagged gate $G^f$, then a following noise-free round of syndrome extraction would correct all these faults. 
The protocol therefore consists of applying the (flagged) unitary gate $G^f$ that has a distinguishable fault set. If any of the detector outcomes $\mathbf{d}^{(0)}$ is non-trivial, a fault has occurred and a non-flagged round of stabilizer measurement is sufficient to correct all errors from fault paths of order $\mathcal{O}(p)$. 
For a trivial detector outcome, we append a round of FT error correction, cf.~Fig.\,\ref{fig:ft_ec}.

Both protocols can be decoded by generating a lookup table for each branch, i.e., by simulating each $\mathcal{O}(p)$ fault path and recording the detector outcomes and propagated errors. The FTEC protocol therefore has two, the FT gate protocol three lookup tables.
In a simulation, we apply these corrections and then verify whether the residual error is still correctable. For that, we again use a lookup table. When constructing this final lookup table, it is important to include all propagated errors in $\mathcal{E}^{(t)}$ that have trivial detector outcomes. Otherwise, errors of weight $>t$ that could be corrected because of distinguishability might be misidentified.

In the following sections, we construct circuits that implement the full logical Clifford group for any LCS code, i.e., the set $\langle \ls[i], \lh[i], \lcx[i][j] \rangle$ for $i,j \in \{1,\dots,k\}$. 
We then introduce flags that lead to distinguishable fault sets for $d=3$ LCS codes such that the gates can be used in the protocols we introduced above. Finally, we benchmark the logical gates in the code family member with the lowest physical qubit count, the $\llbracket 15,3,3 \rrbracket$-LCS code.

\section{Round-robin gates in LCS codes}\label{sec:roundrobin}
In this section, we explain the synthesis of our gates starting from a general decoding-encoding circuit. In the 2-qubit (logical) gate case, this construction leads to the round-robin gates introduced in Refs.\,\cite{jochym2014using,yoder2016universal}. As such, the circuits implementing the gates are not fault-tolerant and require the flag and stabilizer measurements shown in the next Sec.\,\ref{sec:flags}. 
In LCS codes, logical $\lx$- and $\lz$-operators can be chosen to act on the same physical qubits\,\cite{old2024lift}. This implies that the logical operators act on an \emph{odd} number of qubits to ensure anticommutation.
In this section, we focus on the $d=3$ case, for which we construct a fault-tolerant version in the next section. 
We refer to the appendix\,\ref{app:general_logical_ops} for the general (non-FT) construction for arbitrary distance. 

Encoded logical operators can be understood in terms of their action on the stabilizers and logical Pauli operators of the error correcting code: A logical gate $\overline{G}$ transforms the (canonical, i.e.~uniquely anticommuting) logical Pauli operators of logical qubit $i$ in the same way physical Pauli operators are transformed by the physical gate. A logical Hadamard gate, e.g., has to act as $\lh[i] \overline{X}_i \lh[i] = \overline{Z}_i$ and $\lh[i] \overline{Y}_i \lh[i] = -\overline{Y}_i$. Additionally, the logical gate has to preserve the code space, i.e.~keep the stabilizer group invariant, $\overline{G} \mathcal{S} \overline{G}^{\dagger} = \mathcal{S}$.

\begin{figure}
    \centering
    \includegraphics[width=\linewidth]{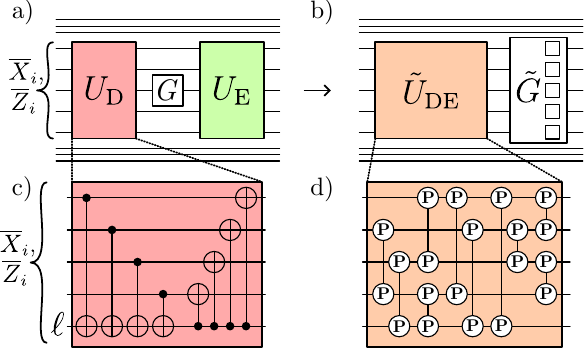}
    \caption{General construction of the round-robin gates. a) A gate $G$ acting on $n_{G}$ logical qubits can be implemented by a unitary decoding circuit $U_D$ acting on the support of the logical operators, mapping them to $n_G$ physical qubits. This is followed by an application of the gate and a re-encoding $U_E = U^{\dagger}_D$. Since $G$ acts on a physical level, this location is considered a bottleneck in the sense that any error on that gate is undetectable and propagates to a logical error through $\mathcal{U}_E$. Here we show a single-qubit gate. b) To circumvent this bottleneck, we commute $G$ through the encoding circuit to the end. This results in an adapted circuit $\tilde{U}_{DE}$ followed by a transformed $\tilde{G}$. If $G$ is a single-qubit Clifford gate, $\tilde{G}$ is transversal. c) A valid choice for a unitary decoding circuit $U_D$ consists of fan-out $\cnot$ gates, mapping the logical $\lx$ and $\lz$ operator onto a single physical qubit $\ell$. d) Commuting single-qubit Clifford gates through the re-encoding circuit results in a $\tilde{U}_{DE}$ that is an all-to-all Pauli-controlled-Pauli gate, i.e.~a $PCP$ gate is applied to every pair of qubits in the support of the logical $\lx$ and $\lz$ operator.}
    \label{fig:gate_transformation}
\end{figure}

We start by constructing circuits that implement a gate $G$ acting on a single logical qubit. 
In a first step, shown in Fig.\,\ref{fig:gate_transformation}\,a), the canonical pair of logical operators $\overline{X}$ and $\overline{Z}$ is contracted onto a physical qubit $\ell$ using a unitary decoding circuit $U_D$. This circuit also has to map all stabilizers such that they have no support on the qubit $\ell$. 
Then, the gate can be applied on the physical level, ensuring the correct transformation of the (now single-qubit) logical Pauli operators, and the invariance of all stabilizers.
The re-encoding circuit $U_E = U^{\dagger}_D$ maps the correctly transformed single-qubit logical Pauli operators back to the full distance logical operators of the code and restores the original stabilizers.

Independent of the actual implementation of $U_D$, this logical gate is not fault-tolerant: a fault in the application of the physical gate is undetectable by construction, since no stabilizer has overlap with the designated qubit $\ell$. This physical error propagates through the re-encoding circuit to a logical error. 
Fault-tolerance can be achieved using concatenation, i.e.~encoding the physical qubits in a code that support a fault-tolerant implementation of the gate $G$\,\cite{yoder2016universal}. This, however, comes at the cost of a large physical qubit overhead.

We now present an alternative path to fault-tolerance, shown in Fig.\,\ref{fig:gate_transformation}\,b), that relies on the structure of the decoding circuit $U_D$. On a high level, we propagate $G$ through the re-encoding circuit to the end of the circuit. This results in an adapted circuit $\tilde{U}_{DE}$ followed by a transformed $\tilde{G}$. 

\begin{figure}
    \centering
    \includegraphics[width=\linewidth]{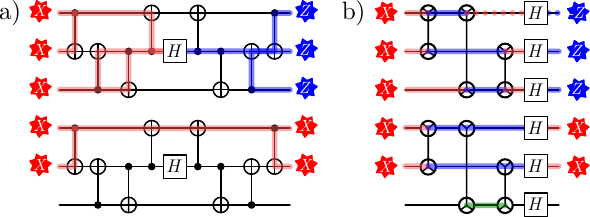}
    \caption{Propagation of Pauli operators through the Hadamard gate on three qubits. We follow the convention $X,Y,Z$ = red, green, blue. A contribution of $-1$ to a global phase is indicated by a dotted highlight. a) In the decoding approach, the logical operators (top) are contracted to the bottleneck location, transformed, and expanded back to the full support. Stabilizer generators with even overlap (bottom) propagate around the bottleneck and stay invariant. b) In the round-robin scheme, the logical operators (top) stay the same through the all-to-all $YCY$s, but pick up a phase of -1 each time two different Pauli operators (here $Z_1 X_3$) propagate through a Pauli-controlled-Pauli (here $Y_1CY_3$) gate. 
    They change their type upon acting with the transversal Hadamard. Stabilizer generators with even overlap (bottom) retain their support after the all-to-all $YCY$s, but change their Pauli type $X \leftrightarrow Z$. The transversal Hadamard fixes this to render the stabilizer group invariant.}
    \label{fig:encoding_propagation}
\end{figure}

Valid  de- and encoding circuits are fan-out $\cnot$ gates on the common support of logical $\lx$- and $\lz$-operators\,\cite{green2001counting},
\begin{align}
    U_D = \bigotimes_{L \in \supp_{G} }\prod_{q \in \supp_{L} \setminus \ell} \cnot[c][\ell] \prod_{q \in \supp_{L} \setminus \ell} \cnot[\ell][c],
\end{align} 
shown in Fig.\,\ref{fig:gate_transformation}c). One can verify that these indeed fulfill the conditions spelled out above: The full overlap logical operators propagate to the designated qubit $\ell$. 
Since stabilizers commute with the logical operators, they have even overlap with the fan-out gates and have no support on qubit $\ell$ after the gate.
We show this circuit with the propagation of Pauli operators for a logical $\lh$ gate in a $d=3$ code in Fig.\,\ref{fig:encoding_propagation}\,a). 
Next, we propagate the central bottleneck single-qubit Clifford gate $G$ to the end of the circuit. 
Applying circuit identities, we obtain circuits $\tilde{U}_{DE}$ that consist of all-to-all \emph{Pauli-controlled-Pauli} gates ($\apcp$), followed by a transversal $\tilde{G} = G^{\otimes n_G}$, shown in Fig.\,\ref{fig:gate_transformation}\,b) and d).
We denote the $\apcp$ gates as
\begin{align}
   PCP'_{Q} := \prod_{i,j \in \{Q\}} P_iCP'_{j}.
\end{align}
with physical Pauli-controlled-Pauli gates that are defined as $PCP' = \frac{I-P}{2} \otimes I + \frac{I+P}{2} \otimes P'$.
In the following circuits, we draw the $PCP$, with $P \in \{X, Y, Z\}$ as
\begin{align}
    XCX : \raisebox{-1.25em}{\includegraphics[height=3em]{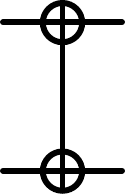}},~~~~~~~
    YCY : \raisebox{-1.25em}{\includegraphics[height=3em]{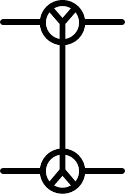}},~~~~~~~
    ZCZ : \raisebox{-1.25em}{\includegraphics[height=3em]{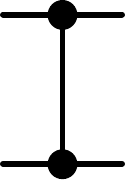}},
\end{align}
which should not be confused with notation for Mølmer-Sørensen gates of Ref.\,\cite{heussen2023strategies}. We also sometimes drop the $Z$ for the control qubit and write $CP \equiv ZCP$.

\begin{figure*}
    \centering
    \includegraphics[width=\linewidth]{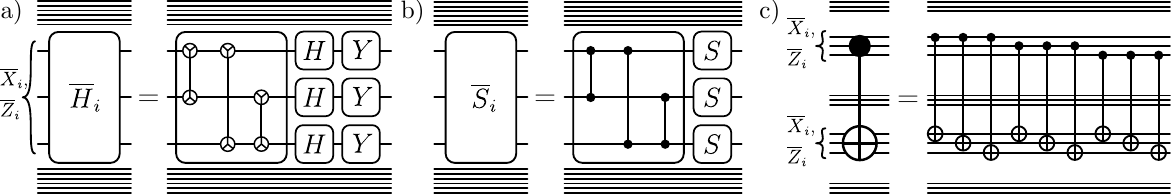}
    \caption{Targeted logical Clifford Gates in $d=3$ LCS codes. a) The logical $\lh$ gate consists of all-to-all $YCY$-gates followed by transversal $H$ and $Y$-gates on the support of the logical operator. b) The logical $\ls$ gate uses all-to-all $ZCZ$-gates followed by transversal $S$-gates on the support of the logical operator. c) The logical intrablock $\lcnot[i][j]$ gate is a round-robin, i.e., all-control-to-all-target $\cnot$.}
    \label{fig:fig_clifford_gates}
\end{figure*}

Similar to before, we now investigate the action of the Clifford $\apcp$ gates on Pauli operators with even (stabilizers) or full (logical operators) overlap.
Note that the $\apcp$ gates act on an odd number of qubits $\{Q\}$, $\abs{\{Q\}} = 1\mod 2$.
Pauli operators that have even overlap with $\{Q\}$ and anticommute  with the controls and targets $P$, change their Pauli types,
\begin{align}
   XCX_{Q} Y^{\otimes 2k} XCX_{Q}  &= Z^{\otimes 2k},\\
    YCY_{Q} X^{\otimes 2k} YCY_{Q} &= Z^{\otimes 2k},\\
    ZCZ_{Q} X^{\otimes 2k} ZCZ_{Q} &= Y^{\otimes 2k}.
\end{align}
On operators with full overlap, the $\apcp$ gates act proportionally to the identity,
\begin{align}
   PCP_{Q} \tilde{P}^{\{Q\}} PCP_{Q}  = (-1)^{\frac{\abs{\{Q\}} - 1 }{2}} \tilde{P}^{\{Q\}},\\
    \qfor P \in \{X,Y,Z\}, \tilde{P} \in \{X,Y,Z\} \setminus P.
\end{align}
We can now combine these actions with transversal gates to generate the full single-qubit logical Clifford group: Consider e.g.~the $YCY_Q$ gate that swaps $X \leftrightarrow Z$ on stabilizers overlapping the logical operators. A transversal Hadamard $H_Q$ fixes that, $Z \leftrightarrow X$ such that the stabilizer group is invariant. The logical operators undergo $\lx \xrightarrow[]{YCY_Q} (-1)^{\frac{\abs{\{Q\}} - 1 }{2}} \lx \xrightarrow[]{H_Q}  (-1)^{\frac{\abs{\{Q\}} - 1 }{2}} \lz  $ and equivalently for $\lx$. A transversal $Y_Q$  if $\frac{\abs{\{Q\}} - 1 }{2} = 1 \bmod 2$ fixes the final phase.
We again show this for a Hadamard gate in a $d=3$ code, together with the propagation of Pauli operators in Fig.\,\ref{fig:encoding_propagation}\,b).
Together with the targeted $\ls$ gate ($ZCZ$-gates and transversal $S_{\supp(L)}$) in Fig.\,\ref{fig:fig_clifford_gates}, this generates the single-qubit Clifford group. 

For logical operators of weight $d$, the decoding approach requires $4(d-1)$ two-qubit gates resulting in a circuit depth $4d-3$. 
The round-robin approach using the $\apcp$ gates needs more, i.e.~$\frac{d(d-1)}{2}$ two-qubit gates. Because the $\apcp$ gates can be parallelized, the circuit depth is reduced to $d+1$. 
With respect to fault-tolerance, the round-robin construction shows another advantage: the central gate location which could already lead to a logical error in $\mathcal{O}(p)$ is removed. This opens the possibility to introduce flags to turn the circuit FT.

Before showing these flag construction, let us add the final ingredient to generate the $n$-logical qubit Clifford group: a targeted two-qubit \emph{intra}-block entangling Clifford gate. 
We follow the same approach as for single-qubit gates: We decode two logical qubits onto two physical qubits $\ell_1, \ell_2$. We apply the desired entangling gates, e.g.~$\cnot[\ell_1][\ell_2]$. Then, both qubits are re-encoded into the code. 
Propagating the entangling gate $G$ through the encoding circuit results in the well-known round-robin gate of Ref.\,\cite{yoder2016universal}. We show the targeted logical $\lcnot$ in Fig.\,\ref{fig:fig_clifford_gates}\,c). In the same way, arbitrary logical $PCP$ can be constructed. Note that the \emph{inter}-block $\lcnot$ between two code blocks is transversal since LCS codes are CSS\,\cite{old2024lift}. This is, however, not a targeted $\lcnot$ but acts on \emph{all} logical qubits of each block simultaneously.

We stress, again, that the round-robin gate constructions get rid of the bottleneck in the middle of the decode-gate-encode approach. There are still, however, single faults that lead to an uncorrectable error in first order. Consider, e.g., an $XX$-fault on the two qubits of the last $YCY$ gate in the logical $\lh$-gate depicted in Fig.\,\ref{fig:fig_flagging_h_gate}\,a). It  propagates to an $IZZ$ error at the end of the circuit. This is not correctable because the fault is indistinguishable from a single-qubit $X$-fault on the first qubit: They both flip the same detectors, and are not stabilizer equivalent but multiply to the logical operator. 
In the next section, we show how to carefully construct stabilizer- and flag measurements to make these logical gate circuits fault-tolerant.

\section{Flag-FT gate constructions}\label{sec:flags}

In what follows, we present our detailed constructions of the FT universal gate set Clifford+$T$ on $\llbracket 15,3,3 \rrbracket$-LCS codes. We specify stabilizer generators and a logical operator basis in appendix\,\ref{app:LCScodes}.
We first consider making single-qubit gates FT. Together with our construction of the FT $CX$ gate, which we discuss subsequently, the FT Clifford group is generated. Finally, we provide two FT magic state preparation schemes that allow us to complete the FT universal gate set via gate teleportation. 

\subsection{FT single-qubit gates}
As seen in the previous section, the round-robin gates are not fault-tolerant. This implies that there are still pairs of locations that lead to undetectable logical errors. Using the $\lh$-gate as a guiding example, we explain how flag constructions based on Ref.\,\cite{chao2018fault} render the fault set $\mathcal{F}^{(1)}$ distinguishable. The flagged gate can then be used in the Flag-FT protocol described above.

The general strategy is to first flag all two qubit gates to catch correlated errors like the $XX$-fault described above. Then, all remaining problematic faults are equivalent to single-qubit faults before the gates. These are detected by inserting stabilizer measurements.

\begin{figure*}
    \centering
    \includegraphics[width=\linewidth]{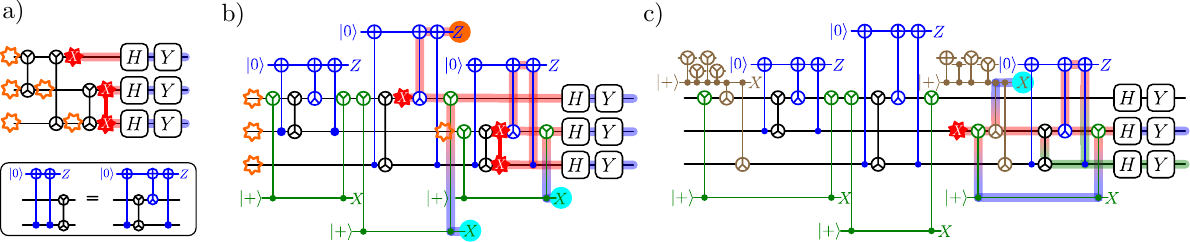}
    \caption{Flagging the single-qubit logical $H$-gate to ensure a distinguishable order $\mathcal{O}(p)$ fault-set $\mathcal{F}^{(1)}$. 
    a) The unflagged $\lh$-gate has numerous fault pairs that propagate to an undetected logical error. They occur at almost any location, indicated by orange stars. We also draw such a fault, a combination of a correlated $XX$-fault after the last $YCY$ gate and a $X$-fault on the first qubit. 
    b) We wrap $Y$- and $Z$-flags around all $YCY$ gates. Since the $Z$- flag does not commute with the $YCY$ gate, propagating it through results in an $XCY$ gate, reflecting the propagating of a Pauli-$Z$ through a $YCY$ gate. We show this in the lower left corner box. 
    This catches all correlated faults on these gates. In particular, the fault pair from the previous example is now detected. The remaining fault-pairs that lead to undetectable logical errors always include fault that are equivalent to faults on the input. We draw  exemplary such locations by orange stars. 
    c) The remaining faults handled by inserting FT stabilizer measurements after the first gate of $Y$ flags (brown). The stabilizers to measure are chosen such that they overlap with the logical operator. We draw an $X$-fault that has previously been undetected but is now caught by the stabilizer measurement. }
    \label{fig:fig_flagging_h_gate}
\end{figure*}

For the bare logical gate, we find via computer-aided inspection that $66$ out of $1953$ possible order-$p^2$ faults lead to an undetected logical error. 48 of these contain correlated 2-qubit faults, like the $XX$ fault of Fig.\,\ref{fig:fig_flagging_h_gate}\,a), that are not equivalent to single-qubit faults before the gate (i.e.~$XI, YI, ZI, IX, IY, IZ, XZ, ZX, XY, YX$). 

All these correlated two-qubit faults on data qubits after a $YCY$-gate can be caught using two auxiliary qubits as flags, cf. Ref.\,\cite{chao2018fault}: for the logical Hadamard gate, the $YCY$ gates are flagged using a $Y$-flag on one qubit i.e.~a flag implementing the measurement of identity using $CY$-gates. This $Y$-flag catches $X$- and $Z$-faults. On the other qubit, a $Z$-flag catches $YY$-faults. We show the same fault-pair from before, that is now detected by flag measurements in Fig.\,\ref{fig:fig_flagging_h_gate}\,b).

There are still faults that are undetected by flags. 
Despite the much larger number of fault locations ($44850$ fault pairs), still only $68$ of those lead to an undetectable logical error. 
These fault pairs contain terms at circuit locations at or equivalent to $X$- or $Y$- faults at the input of the circuit. We show corresponding locations in Fig.\,\ref{fig:fig_flagging_h_gate}\,b).
The reason for these faults not being detected is that they occur before the flags and therefore propagate onto the ancilla twice.
The only chance to identify these remaining faults is to insert stabilizer measurements after the first gate of the initial flags. 
The stabilizers have to be chosen such that they overlap and anticommute with these initial faults. For the $\lh$-gate, the faults are Pauli $X$ and $Z$ - the stabilizer operators therefore have to have a Pauli-$Y$ term at these locations. 
For the logical $\overline{H}_0$, on the logical operator supported on qubits $0,10$ and $12$, we choose to measure the stabilizer $S^{(0)} = \prod_{i=0}^{5} S_X^{(i)}S_Z^{(i)} = Y_0 Y_1 Y_2 Y_{12} Y_{13} Y_{14}$ after the first gate of the $Y$-flag for gate $Y_0C Y_{10}$, and $S^{(1)} = S_X^{(3)}S_Z^{(3)} = X_3 Z_6 Y_9 Y_{10} Y_{12} Y_{14}$ after the first gate of the $Y$-flag for gate $Y_{10}CY_{12}$.
We show this final circuit with a distinguishable fault set in Fig.\,\ref{fig:fig_flagging_h_gate}\,c). 
We show the full circuit for the logical $\ls$ in the appendix\,\ref{app:circuits}
and tabulate the measured stabilizers in Tab.\,\ref{tab:stab_meas} of appendix\,\ref{app:stab_meas}.

\subsection{FT CX gate}

\begin{figure}
    \centering
    \includegraphics[width=0.9\linewidth]{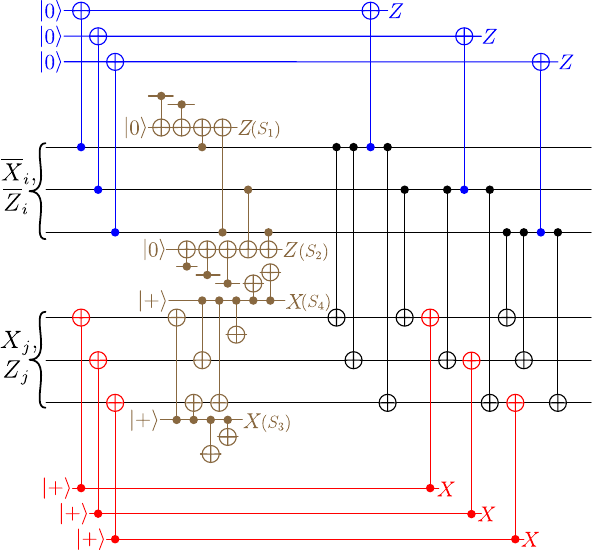}
    \caption{Flagging the two-qubit logical $\lcx$-gate (black) to ensure a distinguishable order $\mathcal{O}(p)$ fault-set $\mathcal{F}^{(1)}$. 
    This can be achieved by wrapping the first two controls and targets of each physical qubit of the round-robin $\lcx$ with $X$- and $Z$-flags (red and blue). To catch the remaining incoming faults, two $Z$- and $X$-stabilizer measurements (brown) are inserted after each flag's first gate.}
    \label{fig:fig_flagged_cnot_gate}
\end{figure}

For the logical $\lcnot$ gate in the bare round-robin construction, 28 out of 16290 fault pairs lead to an undetected logical error. By flagging physical control and target qubits, 36 out of 69378. Note that because the $\lcnot$ has a $Z$-control and an $X$-target, the $Z$- and $X$-flags contain no extra gate due to anti-commutation.  
Also, we find that we don't have to flag each gate separately, but one flag spanning the first two controls and targets of each qubit is sufficient. 
Again, including stabilizer measurements allows one to distinguish the remaining, incoming faults such that $\mathcal{F}^{(1)}$ is distinguishable. 
For a $\lcnot[0][1]$, on the logical operators supported on qubits $(0,10,12)$ and $(1,11,13)$ respectively, the measured stabilizers are $S^{(0)}_Z = Z_0 Z_3 Z_4 Z_{12}$, $S^{(3)}_Z = Z_6 Z_9 Z_{10} Z_{12} Z_{14}$ and $S^{(1)}_X = X_1 X_7 X_8 X_{13}$, $S^{(5)}_X = X_5 X_9 X_{11} X_{13} X_{14}$.
We show the full circuit in Fig.\,\ref{fig:fig_flagged_cnot_gate}. Circuits for other combinations of logical controls and targets have the identical structure while measuring different stabilizer generators. We tabulate those in Tab.\,\ref{tab:stab_meas} in appendix\,\ref{app:stab_meas}. 

The total overhead of the FT logical gates depends on their embedding within an FT protocol. In particular, within the gadget-FT protocols of Sec.\,\ref{sec:notionsft} the number of gates depends on the executed circuit branch. As a reference, we consider the FT flagged gate and enough stabilizer measurements to ensure a generating set has been measured during the gate. We summarize the number of entangling gates and flag qubits required for the single- and two-logical qubit gates in Tab.\,\ref{tab:overheads}. Notably, all logical gates use the same number of flag qubits. The larger number of entangling gates during the logical gate in a $\lcx$ is offset by the smaller number required for flags. This results in very similar numbers for logical single- and two-qubit Clifford gates. 

\begin{table}
    \centering
    \begin{tabular}{l|ccc|c}
    \toprule
     & \multicolumn{3}{c|}{$n(\mathrm{entangling\,gates})$} &  \\
    \multirow{-2}{*}{gate}   & bare & flagged(nFT)  & flagged\!\,+\!\,EC(FT) & \multirow{-2}{*}{$n(\mathrm{flag\,qubits})$} \\
    \midrule
        $\lh$         & $3$ & $18$ & $75$ & $6$\\
        $\ls$         & $3$ & $18$ & $72$ & $6$\\
        $\lcnot$      & $9$ & $21$ & $75$ & $6$\\
    \bottomrule
    \end{tabular}
    \caption{Resource requirements for the logical Clifford gates. We show the number of entangling gates for the bare, non-FT gates and the flagged non-FT logical gates (without intermediate stabilizer measurements). There, the single-qubit logical gates require less compared to the two-qubit logical gate. For the FT gates, we include the entangling gates required for the intermediate stabilizer measurements, as well as a set of stabilizers to complete a generating set. The $\lh$ and the $\lcx$ gate now both require $75$ entangling gates. The $\ls$ gate needs three entangling gates less because the stabilizer generators measured have less weight compared to the $\lh$ gate. All flagged protocols employ at most $6$ additional flag qubits.}
    \label{tab:overheads}
\end{table}

\subsection{FT Magic state preparation}

In order to complete the FT universal gate set Clifford+$T$ for the $\llbracket 15,3,3 \rrbracket$ LCS code, we see two possible routes to take. We require either a flag-FT $T$-gate circuit or a routine to fault-tolerantly prepare a magic state which can then facilitate the FT $T$-gate via gate teleportation\,\cite{knill1998resilient,chamberland2019fault, postler2022demonstration}.
Using the same approach as introduced in the previous sections, we arrive at a (non-FT) logical $T$-gate that uses non-Clifford two-qubit gates, shown in appendix\,\ref{app:general_logical_ops}, Fig.\,\ref{fig:general_d_logical_gates}\,d).
These cannot be flagged with conventional methods, placing a roadblock on the first route to a direct, unitary, flag-FT $T$-gate.
We therefore opt for the second route and implement two strategies for FT magic state preparation\,\cite{goto2016minimizing, chamberland2019fault, heussen2023strategies}. We use the fact that 
\begin{align}
    \overline{H}_i\ket{\overline{\psi}_1,\overline{\psi}_2,...,\overline{H}_i, ..., \overline{\psi}_k} &= +1 \ket{\overline{\psi}_1,\overline{\psi}_2,...,\overline{H}_i, ..., \overline{\psi}_k}
\end{align}
for any $i \in \{1, k\}$ and logical states $\overline{\psi}_i$. We slightly abuse notation to write in short form $\overline{H}\ket{\overline{H}} = +1 \ket{\overline{H}}$, with the logical Hadamard operator
\begin{align}
    \overline{H} &= \frac{\overline{X} + \overline{Z}}{\sqrt{2}}
\end{align}
to prepare the $H$-type magic state
\begin{align}
    \ket{\overline{H}} &= \cos{\left(\frac{\pi}{8}\right)}\ket{\overline{0}} + \sin{\left(\frac{\pi}{8}\right)}\ket{\overline{1}}
\end{align}
on the $i$-th logical qubit.
The first approach is a deterministic protocol that repeatedly measures the logical Hadamard operator on one of the logical qubits. Thereby, a noisy magic state is projected onto the $+1$ eigenstate of $\overline{H}$\,\cite{chamberland2019fault}. A repeat-until-success variant of this protocol allows one to only measure the logical Hadamard operator once, followed by a QEC cycle for detection of residual, otherwise uncorrectable errors. In case that all physical measurements yield $+1$, the magic state preparation succeeds and is guaranteed to be fault-tolerant. The state is discarded in case any measurement yields $-1$ and the state preparation protocol is repeated\,\cite{goto2016minimizing}. A quantitative comparison of these two variants has previously been given with a focus on the Steane code in Ref.\,\cite{heussen2023strategies}. 

Here, we realize \emph{addressable} FT magic state preparation within one LCS codeblock and retain the flexibility to choose either the deterministic or repeat-until-success variant for practical implementation. 
Therefore, every physical gate in the logical Hadamard circuit of Fig.\,\ref{fig:fig_clifford_gates}a) is equipped with an additional control that connects to a measurement qubit. 
Only one flag qubit is required to detect faults that can otherwise propagate to undetectable errors. The full FT circuit construction is given on the physical-qubit level in appendix\,\ref{app:magic}.

\section{Numerical performance analysis}\label{sec:numerics}

We conduct an extensive numerical study on the performance of our FT gates. A single-parameter circuit-level depolarizing noise model is employed, i.e., every noisy operation acting on $n$ qubits is modeled by the ideal version followed by an $n$-qubit depolarizing channel of strength $p$,
\begin{align}
    \mathcal{E}_n(\rho) = (1-p) \rho + \frac{p}{4^n - 1} \sum_{i=1}^{4^n-1} P_n^{(i)} \rho P_n^{(i)} \label{eq:depol} \\
    \quad\mathrm{with}\quad P_n^{(i)} \in \{\{I, X, Y, Z\}^{n} \setminus I^{\otimes n}\}. \nonumber
\end{align}
We use the same depolarizing noise channel after (before) initializations (measurements) of qubits and assume idling locations as noise-free.

We simulate the two approaches discussed in Sec.~\ref{sec:notionsft}: the algorithmically fault-tolerant circuits and the conventionally fault-tolerant protocols.

\subsection{Algorithmic fault-tolerance -- Memory Experiments}
We simulate the Clifford circuits using \texttt{stim}\,\cite{gidney2021stim}. 
For the algorithmic fault-tolerant protocol, we initialize the logical Pauli state $\ket{\overline{000}}$ of the $\llbracket15,3,3\rrbracket$ LCS code without noise and then apply noise-free logical gates to prepare arbitrary logical Pauli eigenstates. This is to focus on the performance of the gates, not including state preparation and measurement (SPAM) errors\footnote{Note that an FT state preparation can be achieved using rounds of (noisy) stabilizer measurement, or flag verification methods as e.g.~shown in Refs.\,\cite{goto2016minimizing, peham2025automated, schmid2025deterministic, zen2025quantum}.}.  For a logical single-qubit gate on the first logical qubit, these are the $6$ states $\ket{\overline{0}}, \ket{\overline{1}}, \ket{\overline{+}}, \ket{\overline{-}},\ket{\overline{+i}}, \ket{\overline{-i}}$. For a logical two-qubit gate, we prepare all $36$ combinations of these on two logical qubits. We choose to prepare the logical qubit(s) untouched by the logical gate in $\ket{\overline{0}}$.
We add a layer of single-qubit depolarizing noise channels, followed by the flagged gate circuits, cf.~Fig.\,\ref{fig:fig_flagging_h_gate}\,c),~Fig.\,\ref{fig:fig_flagged_cnot_gate} and Fig.\,\ref{fig:fig_flagged_s_gate}. We append one round of noisy stabilizer measurements -- the expected number of QEC rounds required per logical gate for algorithmic fault tolerance\,\cite{zhou2025low}. 
In this round, we do not need to measure a full generating set of stabilizers and we omit measuring those that have already been measured during the execution of the gate. If we measured a product of stabilizers (e.g.~$S_X^{(3)}S_Z^{(3)}$ for the logical $\lh$ gate), we omit one of the factors (e.g.~$S_Z^{(3)}$). This ensures that a complete set of stabilizer generators is measured during the protocol. 
We finally measure all stabilizer generators and adequate logical operators without noise. 
For example for $\lh[i]$, if the logical qubit is prepared in an eigenstate of logical $\lx[i]$, we measure $\lz[i]$. For the logical $\lcx$, if the control is a $\lx$-eigenstate, and target in $\lz$, (e.g.~$\ket{\overline{+0}}$) we measure control and target jointly in $\overline{XX}$ and $\overline{ZZ}$. We tabulate all combinations of input bases and logical operators measured in appendix\,\ref{app:stab_meas} in Tab.\,\ref{tab:log_meas}.
We annotate all flag- and stabilizer measurement as \emph{detectors} and the measurements of logical operators as \emph{observables}. 
This allows us to verify the fault-tolerance of these circuits by calling \texttt{.search\_for\_undetectable\_logical\_errors(...)} on the noisy \texttt{stim} circuit, which returns the circuit distance $d_{\mathrm{circ}} = 3$ for all circuits. This distance is the number of elementary faults required to flip an  observable, without flipping any detector.
When sampling, we get a set of \emph{detector} and \emph{observable} measurement outcomes. 
We decode the detector vector using the \texttt{tesseract} decoder\,\cite{beni2025tesseract}, which returns a prediction for the values of the observables. We then count a logical error if the predicted values do not match the measured values of the observable/logical operators.

For the results, we average over all input states with corresponding logical measurements as described above, as well as over all possible logical (control and) target qubits.
As a reference, we also simulate one round of noisy error-correction without any logical gates ($\overline{EC}$).
We show the results in Fig.\,\ref{fig:plot_results}\,a). As an additional FT-check, we fit a function $p_L = a p^\nu$ to the $5$ lowest data points. All logical error rates show a scaling in accordance with the expected fault-tolerant scaling $p_L \propto p^2$.

To estimate a pseudothreshold, i.e.~a point where the logical gate starts to outperform the physical gate, we assume that each of the three physical qubits fails with probability $p_i, i \in [1,2,3]$. The total failure probability that any qubit fails then is $p_{\mathrm{fail}} = 1-(1-p_1)(1-p_2)(1-p_3)$. 
In our general single-parameter noise model, we consider an upper ($p^{(+)}_{\mathrm{th}}$) and lower limit ($p^{(-)}_{\mathrm{th}}$): first, we assume that any of three physical qubits fails with probability $p$, such that $p_{\mathrm{fail}} = 1-(1-p)^3$.
Since in practice, idling qubits fail with some factor $p_{\mathrm{idle}} = \lambda p$, we also look at the limit $\lambda \to 0$ where $p_{\mathrm{fail}} = 1-(1-p)^{n_g}$.

We obtain the pseudothresholds with uncertainties using linear interpolation of the data points upper and lower bounded by the respective statistical errors. 
The upper limits are in the range of  $p^{(+)}_{\mathrm{th}} \approx 4.8\cdot10^{-3}-1.02\cdot10^{-2}$. We summarize the fit parameters in Tab.\,\ref{tab:prefactors} and the pseudothresholds in Tab.\,\ref{tab:pseudothresholds}.

\subsection{Conventional gadget-FT Clifford gate simulations}

\begin{figure*}
    \centering
    \includegraphics[width=0.48\linewidth]{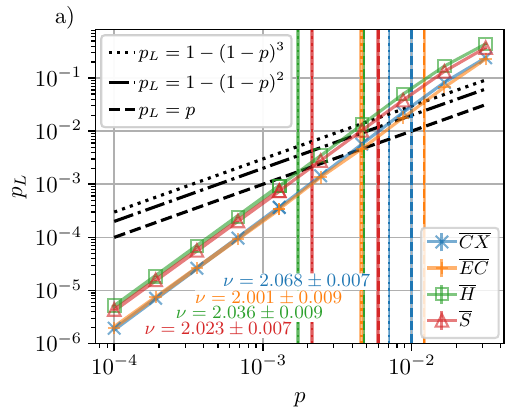}
    \includegraphics[width=0.48\linewidth]{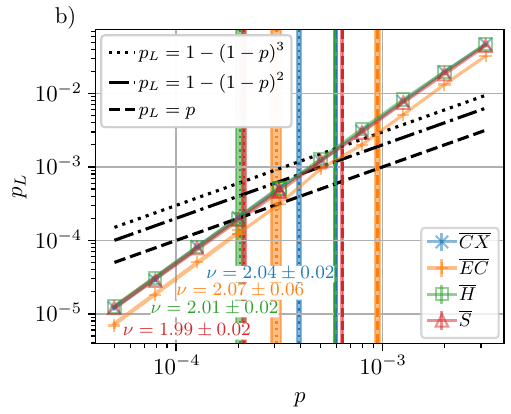}
    \caption{
    a) Flag-FT logical Clifford gates in the algorithmic fault-tolerance protocol. The logical error rate of this protocol depends on the input state and corresponding final logical measurement. We show the average logical error probability over all combinations of input states and logical operators addressed. All logical gates, $\lh$, $\ls$ and $\lcx$, show fault-tolerant scaling ($\nu \approx 2$ for a least-squares fit to $p_L = a p^\nu$). We show lower and upper bounds on pseudothresholds as described in the main text. The breakeven point compared to the failure rate of three physical qubits are in the range of $p^{(+)}_{\mathrm{th}} \approx 4-6.5 \cdot 10^{-3}$. As reference, we show the logical error rate of one round of noisy stabilizer measurements.
    b) Flag-FT logical gate gadgets within the conventional fault-tolerance protocol. This protocol is independent of the input state. We show the average logical error probability over all combinations of logical operators addressed. All logical gates, $\lh$, $\ls$ and $\lcx$, show fault-tolerant scaling ($\nu \approx 2$ for a least-squares fit to $p_L = a p^\nu$). We show lower and upper bounds on pseudothresholds as described in the main text. The breakeven point compared to the failure rate of three physical qubits are in the range of $p^{(+)}_{\mathrm{th}} \approx 5.8-7 \cdot 10^{-4}$. All error bars are standard Monte-Carlo error bars and can be smaller than the line-width. Pseudothresholds are obtained from the intersection of a linear interpolation of data points, as described in the main text and detailed in appendix\,\ref{app:add_sim}. As reference, we show the logical error rate of the FT EC protocol (cf.\,Fig.\,\ref{fig:ft_ec}).}
    \label{fig:plot_results}
\end{figure*}

The conventional gadget fault-tolerant protocol starts with a noise-free initialization, followed by a layer of single-qubit depolarizing noise channels on the data qubits. We then apply the full, noisy 1-FT gate protocol, cf. Fig.\,\ref{fig:ft_gate}. We decode using lookup tables for each circuit branch as described above. We apply the corrections before simulating a perfect round of error correction that puts the state back to the codespace. A logical error occurs if the residual error anti-commutes with \emph{any} logical operator of the code. 
We show results for logical $\ls$, $\lh$ and $\lcx$ gate, averaged over all possible logical targets in Fig.\,\ref{fig:plot_results}\,b). For reference, we show the logical error rate of the bare error correction ($\overline{EC}$) protocol of Fig.\,\ref{fig:ft_ec}.  We again fit a function $p_L = a p^\nu$ to the data and find good accordance with the expected fault-tolerant scaling $p_L \propto p^2$ for all logical gates. We summarize the fitting parameters in Tab.\,\ref{tab:prefactors} and upper and lower bounds on pseudothresholds as described above in Tab.\,\ref{tab:pseudothresholds}.
For all the gates, the upper bounds lie in the range  $p^{(+)}_{\mathrm{th}} \approx 5.8-7\cdot10^{-4}$. 

\begin{table}
    \centering
    \begin{tabular}{lccc}
    \toprule
    FT type & gate & $a$ & $\nu$ \\
    \midrule
                                   & $\lec$        & $200 \pm 10$ & $1.994 \pm 0.008$ \\
                                   & $\lh$         & $710 \pm 50$ & $2.036 \pm 0.009$ \\
                                   & $\ls$         & $530 \pm 20$ & $2.023 \pm 0.007$ \\
    \multirow{-3}{*}{Alg.}         & $\lcnot$      & $340 \pm 20$  & $2.06 \pm 0.02$  \\
    \midrule
                                   & $\lec$        & $6000 \pm 3000$ & $2.07 \pm 0.06$  \\
                                   & $\lh$         & $5300 \pm 800$  & $2.01 \pm 0.02$  \\
                                   & $\ls$         & $3700 \pm 700$  & $1.97 \pm 0.02$  \\
    \multirow{-3}{*}{Gadget}       & $\lcnot$      & $7000 \pm 1000$ & $2.04 \pm 0.02$  \\
    \bottomrule
    \end{tabular}
    \caption{Fit-parameters of logical gates obtained from a least-squares fit of the lowest $5$ data points to $p_L = a p^\nu$. They show the expected fault-tolerant scaling $\nu \approx 2$.}
    \label{tab:prefactors}
\end{table}

\begin{table}
    \centering
    \begin{tabular}{lccc}
    \toprule
    FT type & gate & $p^{(-)}_\mathrm{th}$ & $p^{(+)}_\mathrm{th}$\\
    \midrule
                                   & $\lec$     & $ (4.6 \pm 0.01) \cdot 10^{-3}$  & $(1.22 \pm 0.02) \cdot 10^{-2}$ \\
                                   & $\lh$      & $ (1.8 \pm 0.01) \cdot 10^{-3}$  & $(4.8 \pm 0.02) \cdot 10^{-3}$ \\
                                   & $\ls$      & $ (2.14 \pm 0.03) \cdot 10^{-3}$ & $(5.99 \pm 0.09) \cdot 10^{-3}$ \\
    \multirow{-3}{*}{Alg.}         & $\lcnot$   & $ (6.71 \pm 0.06) \cdot 10^{-3}$ & $(1.020 \pm 0.007) \cdot 10^{-2}$ \\
    \midrule
                                   & $\lec$     & $ (3.1 \pm 0.2)   \cdot 10^{-4}$ & $(9.5 \pm 0.3) \cdot 10^{-4}$\\
                                   & $\lh$      & $ (1.9 \pm 0.1)   \cdot 10^{-4}$ & $(5.88 \pm 0.09) \cdot 10^{-4}$\\
                                   & $\ls$      & $ (2.1 \pm 0.1)   \cdot 10^{-4}$ & $(6.93 \pm 0.07) \cdot 10^{-4}$\\
    \multirow{-3}{*}{Gadget}       & $\lcnot$   & $ (3.93 \pm 0.08) \cdot 10^{-4}$ & $(5.91 \pm 0.09) \cdot 10^{-4}$\\
    \bottomrule
    \end{tabular}
    \caption{The pseudothresholds, upper bounds $p_L(p^{(+)}_\mathrm{th}) = 1-(1-p_\mathrm{th})^3$, are in the order of $10^{-3}-10^{-2}$ for the algorithmic and $10^{-4}$ for the gadget  fault-tolerance. Lower bounds $p_L(p^{(-)}_\mathrm{th}) = 1-(1-p_\mathrm{th})^{n_G}$ depend on the number of qubits the gates act on.}
    \label{tab:pseudothresholds}
\end{table}

\subsection{FT magic state preparation}

Numerical statevector simulations are performed using a modified version of \texttt{PECOS}\,\cite{pecos}. Here, we use the logical fidelity
\begin{align}
    F_{\overline{H}} &= \frac{1}{2}\left(1 + \left(\frac{\langle \overline{X} \rangle + \langle \overline{Z} \rangle}{\sqrt{2}}\right)\right)
\end{align}
of the logical magic state $\ket{\overline{H}}$ to determine logical failure rates $p_L = 1-F_{\overline{H}}$ by evaluating $\langle \overline{X} \rangle$ and $\langle \overline{Z} \rangle$.

\begin{figure*}
    \centering
    \includegraphics[width=0.48\linewidth]{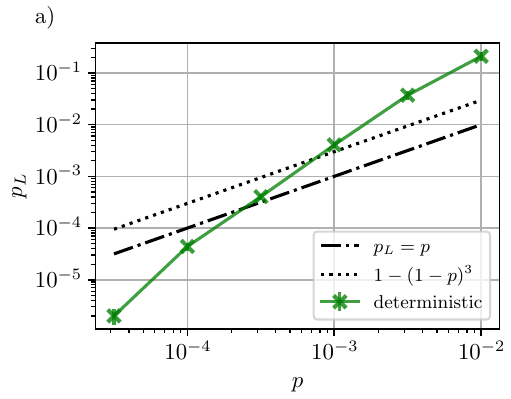}
    \includegraphics[width=0.48\linewidth]{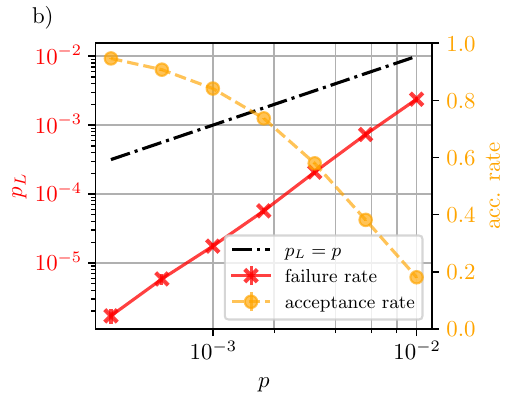}
    \caption{
    a) Failure rate of deterministic FT magic state preparation (green). Breakeven points for outperforming physical qubits (black) with logical magic states lie at similar values as for FT Clifford gates. b) Repeat-until-success FT magic state preparation. While logical failure rates (red) are much lower than physical error rates for all $p$-values of interest, this advantage comes at the price of an exponentially decreasing acceptance rate (orange). Below $p = 10^{-3}$ acceptance rates are feasibly close to unity.}
    \label{fig:t_results}
\end{figure*}

Logical failure rates of (gadget-)FT magic state preparation are showing the expected $p^2$-scaling in Fig.\,\ref{fig:t_results} in the low-$p$ regime. For the deterministic variant, given in Fig.\,\ref{fig:t_results}\,a), we observe a breakeven point as high as approx.~$8 \cdot 10^{-4}$ when comparing to three physical qubits. This point shifts to a slightly smaller value of about $3 \cdot 10^{-4}$ when comparing to a single physical qubit. Both values are very similar to the ones reported above for the gadget-level FT Clifford gates (see Fig.\,\ref{fig:plot_results}\,b)). Remarkably, the analogous curve for the repeat-until-success FT magic state preparation in Fig.\,\ref{fig:t_results}\,b) reveals that an advantage over physical qubits can be expected for the entire range of physical error rates $p \leq 10^{-2}$ considered. However, this advantage comes at the price of increasing the required average number of preparation trials until the state is accepted. While the average acceptance rate at $p = 10^{-2}$ is as low as approx.~$18\%$ (similar to the value reported in Ref.\,\cite{postler2022demonstration}), at the realistically attainable value of $p = 10^{-3}$ already about $85\%$ of trials are accepted. The left-most data point in Fig.\,\ref{fig:t_results}\,b) corresponds to an average acceptance rate of approx.~$95\%$ (see Ref.\,\cite{heussen2023strategies} for an analysis how such values translate to an average expected number of trials until success).

The FT magic state can then be used as an input state to a standard gate teleportation gadget to facilitate the action of a FT $T$-gate on an arbitrary logical target state\,\cite{nielsen2010quantum}. The gate teleportation circuit itself consists of FT Clifford gates only. Yet, there could be different variants to employ: One option would be to prepare the FT magic state on one (or more) of the $k$ logical qubits of one LCS block and teleport the $T$-gate at will to any other qubit within the block or between logical qubits of different code blocks. Alternatively, one may also prepare a full block of $k$ FT magic states in parallel and teleport all at once via a standard transversal $\lcx$ gate to another LCS code block. 

\subsection{Discussion}\label{sec:flaggatediscussion}
In the following, we summarize the resource requirements for the logical gate circuits.
In terms of physical qubit overhead, the flagged single- and two-qubit logical gates on $d=3$ codes require six flag qubits. 
For the single-qubit gates, the flag measurements can be scheduled such that at most two flag qubits are used at the same time, reusing them after measurement and reset.
For the stabilizer measurements, the same qubits used for regular QEC cycles can be used.
In terms of gate overhead, the flag circuits of single-qubit gates employ a total of $15$ additional entangling gates, five per flagged two-qubit gate of the unflagged circuit.
For the two-qubit $\lcx$ gate, only $12$ additional entangling gates are required.
In total, the implementation of all FT Clifford gates and FT QEC cycles require at most $33$ qubits, such that the gates are attractive candidates for near-term implementation in medium-scale devices.

The FT magic state preparation making use of the unflagged logical $\lh$ gate only requires two additional qubits: one acting as measurement qubit, one as a flag for the measurement, cf.\,Fig.\,\ref{fig:fig_flagged_magic}.
However, three-qubit gates are necessary to measure the logical $\lh$ operator. 
Additionally, the deterministic FT magic state preparation needs at most five repetitions. The non-deterministic protocol offers a tradeoff between high-quality magic states and the cost of post-selection.

Next, we put the results of our circuit-level simulations into context. We find pseudothresholds of approximately $2-7 \cdot 10^{-4}$ for the gadget-FT and of approximately $2-6.5 \cdot 10^{-3}$ for the algorithmic FT simulations, cf. Tab.\,\ref{tab:pseudothresholds}.
With respect to the benchmark, i.e.~one round of noisy error correction, we find that performing a flagged gate during one such round, lowers the logical performance only modestly. Most notably, the performance of a $\lcx$ gate in a memory experiment is only slightly worse than that of a bare error correction cycle. In the gadget-FT protocol, all gates show a similar logical error rate. 

Compared to other pseudothresholds reported in the literature, our pseudothresholds are competitive.
Ref.\,\cite{chamberland2016thresholds}, e.g., reports a pseudothreshold of $p_{\mathrm{th}} \approx 4.5\cdot10^{-5}$ for a Hadamard gate in a concatenated $105$-qubit code. 
In Ref.\,\cite{malcolm2025computingefficientlyqldpccodes}, the authors report a logical Clifford gate pseudothreshold of $1.85\cdot10^{-4}$ per round for a logical randomized benchmarking experiment.
The surgery-based protocols of Ref.\,\cite{baspin2025fast} for a $\llbracket 42,6,4 \rrbracket$ multi-cycle qLDPC code have been simulated with a phenomenological noise model, under which they report a pseudothreshold of $p_{\mathrm{th}}^{(-)} \approx 7\cdot10^{-3}$. It is reasonable to expect a worse performance under a circuit-level noise model. 
Ultimately, although not directly comparable because of the different noise models and figures of merit, our pseudothresholds are in similar orders of magnitude, underlining the near-term competitiveness of flag-based protocols.

Next, we ask whether we can expect a finite pseudothreshold scaling these protocols to larger distances. 
Using a fault-tolerant protocol, the logical error rate scales as $p_L = a p^{t+1} + \mathcal{O}(p^{t+2})$ for small physical error rates. 
Here, $a$ is the prefactor that is determined by the number of uncorrectable faults in the lowest (i.e.~$t+1$) order. In general, $a$ depends on the code and the circuit considered. For our $d=3$ code, the fit-parameter $a$ (cf.~Tab.~\ref{tab:prefactors}) therefore is an estimate for that number and is determined by how many $\mathcal{O}(p^2)$ faults lead to a logical failure. 
In the algorithmic fault-tolerance protocol, this number is substantially lower compared to the gadget fault-tolerance. 
This is because the final single-qubit measurements that act as a "noise-free" round of stabilizer-measurement are included in a correlated decoding. 
In particular, this enables more compact protocols:  no branching  is required (cf. flagged gadget protocol) to acquire confidence in the values of the stabilizers. 
The physical failure rate scales as $p_{\mathrm{fail}} = 1 - (1-p)^k \sim kp$ as $p \to 0$. For the LCS codes and assuming $k = d$, it follows
\begin{align}
    p_{\mathrm{th}} = \left( \frac{k(d)}{a(d)} \right)^{\frac{1}{t+1}} = \left( \frac{d}{a(d)} \right)^{\frac{2}{d+1}}.\label{eq:thresha}
\end{align}
Whether the pseudothreshold is finite or goes to zero is determined by the functional dependence of the prefactor $a(d)$ on the distance $d$. If $a$ scales at most exponentially in $d$, that is $a \sim d^{\beta} e^{\alpha d}$, we can see from Eq.~\eqref{eq:thresha} that $p_{\mathrm{th}}$ approaches a non-zero constant value in the limit $d\rightarrow\infty$. Additional "back-of-the-envelope" estimations are provided in appendix\,\ref{sec:scaleflag}.

\section{Conclusions and Outlook}\label{sec:conclusion}

High-rate qLDPC codes have received major attention in the QEC community recently because of their favorable properties as fault-tolerant quantum memories. Yet, performing FT computation directly within the code space is a pending question for many of such codes. 

In this work, we have given an analysis of how to realize the FT universal gate set Clifford+$T$ in lift-connected surface codes. Focusing on near-term implementations, we put the $\llbracket15,3,3\rrbracket$ member of the LCS code family center-stage. We start from a general prescription on how to construct all unitary Clifford gates for \emph{any} member of the code family that is based on the underlying lifted-product structure of the code definition. Then, we employ flag qubit constructions and FT magic state preparation to ensure fault tolerance of our universal gate set against $t=1$ arbitrary Pauli fault. A rigorous analysis of the performance of our circuits under a general depolarizing noise model reveals competitive breakeven points: We find that, for a strict gadget-level fault tolerance definition, an advantage over physical gates can be expected around physical error rates of $10^{-4}$. These values drastically improve to well above $10^{-3}$ when adopting the notion of algorithmic fault tolerance, which allows one to decode the full experiment including the final single-qubit data measurements. 

The obvious way forward is to generalize the flag construction used for $d=3$ Clifford gates to higher-distance round-robin gates. Given that generalizations for error-correction circuits have been devised in Ref.\,\cite{chamberland2018flag}, we expect similar constructions to also work for logical gates. 

Three-qubit-gates, involved in the FT magic state preparation, could be made natively available in state-of-the-art or near-term quantum processors, for instance, based on trapped ions or neutral atoms \cite{espinoza2021highfidelity, evered2023high, sahay2025foldtransversalsurfacecodecultivation, Pecorari_2025,old2025faulttolerantstabilizermeasurementssurface}. Instead decomposing these gates into sequences of single- and two-qubit gates may lead to a decrease in logical fidelity or, in the worst case, increase circuit depth to prohibitive levels for some platforms.
As an alternative to directly realize an FT $T$-gate, we conjecture that specialized flag-like constructions for a unitary $T$-gate circuit that catch both $X$- and $Z$-faults may be conceivable.

\section*{Data availability}
All data shown in this manuscript is available at \href{https://doi.org/10.5281/zenodo.17574057}{https://doi.org/10.5281/zenodo.17574057} (Ref.\,\cite{josias_old_2025_17574057}).

\section*{Code availability}
Software code used in this project is available from the corresponding authors upon reasonable request.

\section*{Author contributions}
JO, JB and SH devised all gate constructions. JO and SH performed numerical simulations and wrote the manuscript with contributions from all authors. MM and SH supervised the project.

\section*{Acknowledgments}

We gratefully acknowledge funding from the German Federal Ministry of Research, Technology and Space (BMFTR) as part of the Research Program Quantum Systems, research project 13N17317 ("SQale"). We furthermore acknowledge support by the European Union’s Horizon Europe research and innovation programme under Grant Agreement No.~101114305 (“MILLENION-SGA1” EU Project). This research is also part of the Munich Quantum Valley (K-8), which is supported by the Bavarian state government with funds from the Hightech Agenda Bayern Plus. We additionally acknowledge support by the BMFTR project MUNIQC-ATOMS (Grant No. 13N16070).
MM and JO acknowledge support by the Deutsche Forschungsgemeinschaft (DFG, German Research Foundation) under Germany’s Excellence Strategy “Cluster of Excellence Matter and Light for Quantum Computing (ML4Q) EXC 2004/1” 390534769. 
The authors gratefully acknowledge the computing time provided to them at the NHR Center NHR4CES at RWTH Aachen University (Project No. p0020074). This is funded by the Federal Ministry of Education and Research and the state governments participating on the basis of the resolutions of the GWK for national high performance computing at universities.

\bibliographystyle{bibstyle}
\bibliography{references}

@article{peham2025automated,
	author = {Peham, Tom and Schmid, Ludwig and Berent, Lucas and M{\ifmmode\ddot{u}\else\"{u}\fi}ller, Markus and Wille, Robert},
	title = {{Automated Synthesis of Fault-Tolerant State Preparation Circuits for Quantum Error-Correction Codes}},
	journal = {PRX Quantum},
	volume = {6},
	number = {2},
	pages = {020330},
	year = {2025},
	month = may,
	publisher = {American Physical Society},
	url = {https://doi.org/10.1103/PRXQuantum.6.020330},
	doi = {10.1103/PRXQuantum.6.020330}
}

@INPROCEEDINGS{schmid2025deterministic,
  author={Schmid, Ludwig and Peham, Tom and Berent, Lucas and Müller, Markus and Wille, Robert},
  booktitle={2025 Design, Automation \& Test in Europe Conference (DATE)}, 
  title={Deterministic Fault-Tolerant State Preparation for Near-Term Quantum Error Correction: Automatic Synthesis Using Boolean Satisfiability}, 
  year={2025},
  volume={},
  number={},
  pages={1-7},
  doi={10.23919/DATE64628.2025.10992896}}

@article{zen2025quantum,
	author = {Zen, Remmy and Olle, Jan and Colmenarez, Luis and Puviani, Matteo and M{\ifmmode\ddot{u}\else\"{u}\fi}ller, Markus and Marquardt, Florian},
	title = {{Quantum Circuit Discovery for Fault-Tolerant Logical State Preparation with Reinforcement Learning}},
	journal = {Phys. Rev. X},
	volume = {15},
	number = {4},
	pages = {041012},
	year = {2025},
	month = oct,
	publisher = {American Physical Society},
	url = {https://doi.org/10.1103/gqpr-dgz7},
	doi = {10.1103/gqpr-dgz7}
}

@misc{liang2025self,
	author = {Liang, Zijian and Chen, Yu-An},
	title = {{Self-dual bivariate bicycle codes with transversal Clifford gates}},
	year = {2025},
	month = oct,
	eprint = {2510.05211},
    archivePrefix={arXiv},
}

@article{cohen2022low,
	author = {Cohen, Lawrence Z. and Kim, Isaac H. and Bartlett, Stephen D. and Brown, Benjamin J.},
	title = {{Low-overhead fault-tolerant quantum computing using long-range connectivity}},
	journal = {Sci. Adv.},
	volume = {8},
	number = {20},
	year = {2022},
	month = may,
	issn = {2375-2548},
	publisher = {American Association for the Advancement of Science},
	url = {https://doi.org/10.1126/sciadv.abn1717},
}

@misc{williamson2024low,
	author = {Williamson, Dominic J. and Yoder, Theodore J.},
	title = {{Low-overhead fault-tolerant quantum computation by gauging logical operators}},
	journal = {arXiv},
	year = {2024},
	month = oct,
	eprint = {2410.02213},
    archivePrefix={arXiv},
}

@misc{cowtan2024ssip,
	author = {Cowtan, Alexander},
	title = {{SSIP: automated surgery with quantum LDPC codes}},
	journal = {arXiv},
	year = {2024},
	month = jul,
	eprint = {2407.09423},
    archivePrefix={arXiv},
}

@article{ide2025fault,
	author = {Ide, Benjamin and Gowda, Manoj G. and Nadkarni, Priya J. and Dauphinais, Guillaume},
	title = {{Fault-Tolerant Logical Measurements via Homological Measurement}},
	journal = {Phys. Rev. X},
	volume = {15},
	number = {2},
	pages = {021088},
	year = {2025},
	month = jun,
	publisher = {American Physical Society},
	url = {https://doi.org/10.1103/PhysRevX.15.021088},
	doi = {10.1103/PhysRevX.15.021088}
}

@article{horsman2012surface,
	author = {Horsman, Dominic and Fowler, Austin G. and Devitt, Simon and Van Meter, Rodney},
	title = {{Surface code quantum computing by lattice surgery}},
	journal = {New J. Phys.},
	volume = {14},
	number = {12},
	pages = {123011},
	year = {2012},
	month = dec,
	issn = {1367-2630},
	publisher = {IOP Publishing},
	url = {https://doi.org/10.1088/1367-2630/14/12/123011},
	doi = {10.1088/1367-2630/14/12/123011}
}

@misc{brun2015teleportation,
	author = {Brun, Todd A. and Zheng, Yi-Cong and Hsu, Kung-Chuan and Job, Joshua and Lai, Ching-Yi},
	title = {{Teleportation-based Fault-tolerant Quantum Computation in Multi-qubit Large Block Codes}},
	journal = {arXiv},
	year = {2015},
	month = apr,
	eprint = {1504.03913},
    archivePrefix={arXiv},
}

@article{breuckmann2024fold,
	author = {Breuckmann, Nikolas P. and Burton, Simon},
	title = {{Fold-Transversal Clifford Gates for Quantum Codes}},
	journal = {Quantum},
	volume = {8},
	pages = {1372},
	year = {2024},
	month = jun,
	publisher = {Verein zur F{\ifmmode\ddot{o}\else\"{o}\fi}rderung des Open Access Publizierens in den Quantenwissenschaften},
	url = {https://doi.org/10.22331/q-2024-06-13-1372},
	doi = {10.22331/q-2024-06-13-1372}
}

@article{bombin2006topological,
	author = {Bombin, H. and Martin-Delgado, M. A.},
	title = {{Topological quantum error correction with optimal encoding rate}},
	journal = {Phys. Rev. A},
	volume = {73},
	number = {6},
	pages = {062303},
	year = {2006},
	month = jun,
	publisher = {American Physical Society},
	url = {https://doi.org/10.1103/PhysRevA.73.062303},
	doi = {10.1103/PhysRevA.73.062303}
}

@article{eastin2009restrictions,
	author = {Eastin, Bryan and Knill, Emanuel},
	title = {{Restrictions on Transversal Encoded Quantum Gate Sets}},
	journal = {Phys. Rev. Lett.},
	volume = {102},
	number = {11},
	pages = {110502},
	year = {2009},
	month = mar,
	publisher = {American Physical Society},
	url = {https://doi.org/10.1103/PhysRevLett.102.110502},
	doi = {10.1103/PhysRevLett.102.110502}
}

@article{divincenzo2007effective,
	author = {DiVincenzo, David P. and Aliferis, Panos},
	title = {{Effective Fault-Tolerant Quantum Computation with Slow Measurements}},
	journal = {Phys. Rev. Lett.},
	volume = {98},
	number = {2},
	pages = {020501},
	year = {2007},
	month = jan,
	publisher = {American Physical Society},
	url = {https://doi.org/10.1103/PhysRevLett.98.020501},
	doi = {10.1103/PhysRevLett.98.020501}
}

@misc{gottesman1997thesis,
	author = {Gottesman, Daniel},
	publisher = {California Institute of Technology},
	title = {Stabilizer codes and quantum error correction},
    eprint = {quant-ph/9705052},
    archivePrefix={arXiv},
	year = {1997}
}

@misc{dasu2025breaking,
	author = {Dasu, Shival and Burton, Simon and Mayer, Karl and Amaro, David and Gerber, Justin A. and Gilmore, Kevin and Gresh, Dan and DelVento, Davide and Potter, Andrew C. and Hayes, David},
	title = {{Breaking even with magic: demonstration of a high-fidelity logical non-Clifford gate}},
	journal = {arXiv},
	year = {2025},
	month = jun,
	eprint = {2506.14688},
    archivePrefix={arXiv},
}

@article{lacroix2025scaling,
	author = {Lacroix, N. and Bourassa, A. and Heras, F. J. H. and Zhang, L. M. and Bausch, J. and Senior, A. W. and Edlich, T. and Shutty, N. and Sivak, V. and Bengtsson, A. and McEwen, M. and Higgott, O. and Kafri, D. and Claes, J. and Morvan, A. and Chen, Z. and Zalcman, A. and Madhuk, S. and Acharya, R. and Aghababaie Beni, L. and Aigeldinger, G. and Alcaraz, R. and Andersen, T. I. and Ansmann, M. and Arute, F. and Arya, K. and Asfaw, A. and Atalaya, J. and Babbush, R. and Ballard, B. and Bardin, J. C. and Bilmes, A. and Blackwell, S. and Bovaird, J. and Bowers, D. and Brill, L. and Broughton, M. and Browne, D. A. and Buchea, B. and Buckley, B. B. and Burger, T. and Burkett, B. and Bushnell, N. and Cabrera, A. and Campero, J. and Chang, H.-S. and Chiaro, B. and Chih, L.-Y. and Cleland, A. Y. and Cogan, J. and Collins, R. and Conner, P. and Courtney, W. and Crook, A. L. and Curtin, B. and Das, S. and Demura, S. and De Lorenzo, L. and Di Paolo, A. and Donohoe, P. and Drozdov, I. and Dunsworth, A. and Eickbusch, A. and Elbag, A. Moshe and Elzouka, M. and Erickson, C. and Ferreira, V. S. and Flores Burgos, L. and Forati, E. and Fowler, A. G. and Foxen, B. and Ganjam, S. and Garcia, G. and Gasca, R. and Genois, {\ifmmode\acute{E}\else\'{E}\fi}. and Giang, W. and Gilboa, D. and Gosula, R. and Grajales Dau, A. and Graumann, D. and Greene, A. and Gross, J. A. and Ha, T. and Habegger, S. and Hansen, M. and Harrigan, M. P. and Harrington, S. D. and Heslin, S. and Heu, P. and Hiltermann, R. and Hilton, J. and Hong, S. and Huang, H.-Y. and Huff, A. and Huggins, W. J. and Jeffrey, E. and Jiang, Z. and Jin, X. and Joshi, C. and Juhas, P. and Kabel, A. and Kang, H. and Karamlou, A. H. and Kechedzhi, K. and Khaire, T. and Khattar, T. and Khezri, M. and Kim, S. and Klimov, P. V. and Kobrin, B. and Korotkov, A. N. and Kostritsa, F. and Kreikebaum, J. Mark and Kurilovich, V. D. and Landhuis, D. and Lange-Dei, T. and Langley, B. W. and Laptev, P. and Lau, K.-M. and Ledford, J. and Lee, K. and Lester, B. J. and Le Guevel, L. and Li, W. Yan and Li, Y. and Lill, A. T. and Livingston, W. P. and Locharla, A. and Lucero, E. and Lundahl, D. and Lunt, A. and Maloney, A. and Mandr{\ifmmode\grave{a}\else\`{a}\fi}, S. and Martin, L. S. and Martin, O. and Maxfield, C. and McClean, J. R. and Meeks, S. and Megrant, A. and Miao, K. C. and Molavi, R. and Molina, S. and Montazeri, S. and Movassagh, R. and Neill, C. and Newman, M. and Nguyen, A. and Nguyen, M. and Ni, C.-H. and Niu, M. Y. and Oas, L. and Oliver, W. D. and Orosco, R. and Ottosson, K. and Pizzuto, A. and Potter, R. and Pritchard, O. and Quintana, C. and Ramachandran, G. and Reagor, M. J. and Resnick, R. and Rhodes, D. M. and Roberts, G. and Rosenberg, E. and Rosenfeld, E. and Rossi, E. and Roushan, P. and Sankaragomathi, K. and Schurkus, H. F. and Shearn, M. J. and Shorter, A. and Shvarts, V. and Small, S. and Smith, W. Clarke and Springer, S. and Sterling, G. and Suchard, J. and Szasz, A. and Sztein, A. and Thor, D. and Tomita, E. and Torres, A. and Torunbalci, M. Mert and Vaishnav, A. and Vargas, J. and Vdovichev, S. and Vidal, G. and Vollgraff Heidweiller, C. and Waltman, S. and Waltz, J. and Wang, S. X. and Ware, B. and Weidel, T. and White, T. and Wong, K. and Woo, B. W. K. and Woodson, M. and Xing, C. and Yao, Z. Jamie and Yeh, P. and Ying, B. and Yoo, J. and Yosri, N. and Young, G. and Zhang, Y. and Zhu, N. and Zobrist, N. and Neven, H. and Kohli, P. and Davies, A. and Boixo, S. and Kelly, J. and Jones, C. and Gidney, C. and Satzinger, K. J.},
	title = {{Scaling and logic in the colour code on a superconducting quantum processor}},
	journal = {Nature},
	volume = {645},
	number = {8081},
	pages = {614--619},
	year = {2025},
	month = sep,
	issn = {1476-4687},
	publisher = {Nature Publishing Group},
	url = {https://doi.org/10.1038/s41586-025-09061-4},
	doi = {10.1038/s41586-025-09061-4}
}

@article{breuckmann2021quantum,
	author = {Breuckmann, Nikolas P. and Eberhardt, Jens Niklas},
	title = {{Quantum Low-Density Parity-Check Codes}},
	journal = {PRX Quantum},
	volume = {2},
	number = {4},
	pages = {040101},
	year = {2021},
	month = oct,
	publisher = {American Physical Society},
	url = {https://doi.org/10.1103/PRXQuantum.2.040101},
	doi = {10.1103/PRXQuantum.2.040101}
}

@inproceedings{panteleev2022asymptotically,
	author = {Panteleev, Pavel and Kalachev, Gleb},
	booktitle = {Proceedings of the 54th Annual ACM SIGACT Symposium on Theory of Computing},
	doi = {10.1145/3519935.3520017},
	pages = {375--388},
	title = {Asymptotically good quantum and locally testable classical LDPC codes},
	url = {https://doi.org/10.1145/3519935.3520017},
	year = {2022}}

@article{breuckmann2021balanced,
	author = {Breuckmann, Nikolas P and Eberhardt, Jens N},
	doi = {10.1109/TIT.2021.3097347},
	journal = {IEEE Transactions on Information Theory},
	number = {10},
	pages = {6653--6674},
	publisher = {IEEE},
	title = {Balanced product quantum codes},
	url = {https://doi.org/10.1109/TIT.2021.3097347},
	volume = {67},
	year = {2021}}

@misc{green2001counting,
	author = {Green, Frederic and Homer, Steven and Moore, Cristopher and Pollett, Christopher},
	title = {{Counting, fanout, and the complexity of quantum ACC}},
	year = {2001},
	month = jun,
	eprint = {quant-ph/0106017},
    archivePrefix={arXiv},
}

@article{yoder2017the,
	author = {Yoder, Theodore J. and Kim, Isaac H.},
	title = {{The surface code with a twist}},
	journal = {Quantum},
	volume = {1},
	pages = {2},
	year = {2017},
	month = apr,
	publisher = {Verein zur F{\ifmmode\ddot{o}\else\"{o}\fi}rderung des Open Access Publizierens in den Quantenwissenschaften},
	url = {https://doi.org/10.22331/q-2017-04-25-2},
	doi = {10.22331/q-2017-04-25-2}
}

@article{chao2018quantum,
	author = {Chao, Rui and Reichardt, Ben W.},
	title = {{Quantum Error Correction with Only Two Extra Qubits}},
	journal = {Phys. Rev. Lett.},
	volume = {121},
	number = {5},
	pages = {050502},
	year = {2018},
	month = aug,
	publisher = {American Physical Society},
	url = {https://doi.org/10.1103/PhysRevLett.121.050502},
	doi = {10.1103/PhysRevLett.121.050502}
}

@misc{xu2025batched,
	author = {Xu, Qian and Zhou, Hengyun and Bluvstein, Dolev and Cain, Madelyn and Kalinowski, Marcin and Preskill, John and Lukin, Mikhail D. and Maskara, Nishad},
	title = {{Batched high-rate logical operations for quantum LDPC codes}},
	year = {2025},
	month = oct,
	eprint = {2510.06159},
    archivePrefix={arXiv},
}

@misc{baspin2025fast,
	author = {Baspin, Nou{\ifmmode\acute{e}\else\'{e}\fi}dyn and Berent, Lucas and Cohen, Lawrence Z.},
	title = {{Fast surgery for quantum LDPC codes}},
	year = {2025},
	month = oct,
	eprint = {2510.04521},
	archivePrefix={arXiv},
}

@article{tansuwannont2022achieving,
	author = {Tansuwannont, Theerapat and Leung, Debbie},
	doi = {10.1103/PRXQuantum.3.030322},
	journal = {PRX Quantum},
	month = aug,
	number = {3},
	pages = {030322},
	publisher = {American Physical Society},
	title = {{Achieving Fault Tolerance on Capped Color Codes with Few Ancillas}},
	url = {https://doi.org/10.1103/PRXQuantum.3.030322},
	volume = {3},
	year = {2022}}

@article{magdalena2025xyz,
	author = {Magdalena de la Fuente, Julio C. and Old, Josias and Townsend-Teague, Alex and Rispler, Manuel and Eisert, Jens and M{\ifmmode\ddot{u}\else\"{u}\fi}ller, Markus},
	doi = {10.1103/PRXQuantum.6.010360},
	journal = {PRX Quantum},
	month = mar,
	number = {1},
	pages = {010360},
	publisher = {American Physical Society},
	title = {{$\mathrm{XYZ}$ Ruby Code: Making a Case for a Three-Colored Graphical Calculus for Quantum Error Correction in Spacetime}},
	url = {https://doi.org/10.1103/PRXQuantum.6.010360},
	volume = {6},
	year = {2025}}

@article{bombin2024unifying,
	author = {Bombin, Hector and Litinski, Daniel and Nickerson, Naomi and Pastawski, Fernando and Roberts, Sam},
	title = {{Unifying flavors of fault tolerance with the ZX calculus}},
	journal = {Quantum},
	volume = {8},
	pages = {1379},
	year = {2024},
	month = jun,
	publisher = {Verein zur F{\ifmmode\ddot{o}\else\"{o}\fi}rderung des Open Access Publizierens in den Quantenwissenschaften},
	url = {https://doi.org/10.22331/q-2024-06-18-1379},
	doi = {10.22331/q-2024-06-18-1379}
}

@article{derks2025designing,
	author = {Derks, Peter-Jan H. S. and Townsend-Teague, Alex and Burchards, Ansgar G. and Eisert, Jens},
	title = {{Designing fault-tolerant circuits using detector error models}},
	journal = {Quantum},
	volume = {9},
	pages = {1905},
	year = {2025},
	month = nov,
	publisher = {Verein zur F{\ifmmode\ddot{o}\else\"{o}\fi}rderung des Open Access Publizierens in den Quantenwissenschaften},
	url = {https://doi.org/10.22331/q-2025-11-06-1905},
	doi = {10.22331/q-2025-11-06-1905}
}

@misc{delfosse2023spacetime,
	author = {Delfosse, Nicolas and Paetznick, Adam},
	title = {{Spacetime codes of Clifford circuits}},
	year = {2023},
	month = apr,
	eprint = {2304.05943},
	archivePrefix={arXiv},
}

@article{dennis2002topological,
	author = {Dennis, Eric and Kitaev, Alexei and Landahl, Andrew and Preskill, John},
	doi = {10.1063/1.1499754},
	issn = {0022-2488},
	journal = {J. Math. Phys.},
	month = sep,
	number = {9},
	pages = {4452--4505},
	publisher = {AIP Publishing},
	title = {{Topological quantum memory}},
	url = {https://doi.org/10.1063/1.1499754},
	volume = {43},
	year = {2002}}

@inproceedings{shor1996fault,
	author = {Shor, P.W.},
	title = {Fault-tolerant quantum computation},
	booktitle = {Proceedings of 37th Conference on Foundations of Computer Science},
	doi = {10.1109/SFCS.1996.548464},
	pages = {56-65},
	year = {1996}}

@article{steane1997active,
	author = {Steane, A. M.},
	doi = {10.1103/PhysRevLett.78.2252},
	journal = {Phys. Rev. Lett.},
	month = mar,
	number = {11},
	pages = {2252--2255},
	publisher = {American Physical Society},
	title = {{Active Stabilization, Quantum Computation, and Quantum State Synthesis}},
	url = {https://doi.org/10.1103/PhysRevLett.78.2252},
	volume = {78},
	year = {1997}}

@article{quintavalle2023partitioning,
	author = {Quintavalle, Armanda O. and Webster, Paul and Vasmer, Michael},
	title = {{Partitioning qubits in hypergraph product codes to implement logical gates}},
	journal = {Quantum},
	volume = {7},
	pages = {1153},
	year = {2023},
	month = oct,
	publisher = {Verein zur F{\ifmmode\ddot{o}\else\"{o}\fi}rderung des Open Access Publizierens in den Quantenwissenschaften},
	url = {https://doi.org/10.22331/q-2023-10-24-1153},
	doi = {10.22331/q-2023-10-24-1153}
}

@article{sayginel2025fault,
	author = {Sayginel, Hasan and Koutsioumpas, Stergios and Webster, Mark and Rajput, Abhishek and Browne, Dan E.},
	title = {{Fault-Tolerant Logical Clifford Gates from Code Automorphisms}},
	journal = {PRX Quantum},
	volume = {6},
	number = {3},
	pages = {030343},
	year = {2025},
	month = sep,
	publisher = {American Physical Society},
	url = {https://doi.org/10.1103/vf7v-cpq9},
	doi = {10.1103/vf7v-cpq9}
}

@article{daguerre2025experimental,
	author = {Daguerre, Lucas and Blume-Kohout, Robin and Brown, Natalie C. and Hayes, David and Kim, Isaac H.},
	title = {{Experimental Demonstration of High-Fidelity Logical Magic States from Code Switching}},
	journal = {Phys. Rev. X},
	volume = {15},
	number = {4},
	pages = {041008},
	year = {2025},
	month = oct,
	publisher = {American Physical Society},
	url = {https://doi.org/10.1103/dck4-x9c2},
	doi = {10.1103/dck4-x9c2}
}

@article{sales2025experimental,
	author = {Sales Rodriguez, Pedro and Robinson, John M. and Jepsen, Paul Niklas and He, Zhiyang and Duckering, Casey and Zhao, Chen and Wu, Kai-Hsin and Campo, Joseph and Bagnall, Kevin and Kwon, Minho and Karolyshyn, Thomas and Weinberg, Phillip and Cain, Madelyn and Evered, Simon J. and Geim, Alexandra A. and Kalinowski, Marcin and Li, Sophie H. and Manovitz, Tom and Amato-Grill, Jesse and Basham, James I. and Bernstein, Liane and Braverman, Boris and Bylinskii, Alexei and Choukri, Adam and DeAngelo, Robert J. and Fang, Fang and Fieweger, Connor and Frederick, Paige and Haines, David and Hamdan, Majd and Hammett, Julian and Hsu, Ning and Hu, Ming-Guang and Huber, Florian and Jia, Ningyuan and Kedar, Dhruv and Kornja{\ifmmode\check{c}\else\v{c}\fi}a, Milan and Liu, Fangli and Long, John and Lopatin, Jonathan and Lopes, Pedro L. S. and Luo, Xiu-Zhe and Macr{\ifmmode\grave{\imath}\else\`{\i}\fi}, Tommaso and Markovi{\ifmmode\acute{c}\else\'{c}\fi}, Ognjen and Mart{\ifmmode\acute{\imath}\else\'{\i}\fi}nez-Mart{\ifmmode\acute{\imath}\else\'{\i}\fi}nez, Luis A. and Meng, Xianmei and Ostermann, Stefan and Ostroumov, Evgeny and Paquette, David and Qiang, Zexuan and Shofman, Vadim and Singh, Anshuman and Singh, Manuj and Sinha, Nandan and Thoreen, Henry and Wan, Noel and Wang, Yiping and Waxman-Lenz, Daniel and Wong, Tak and Wurtz, Jonathan and Zhdanov, Andrii and Zheng, Laurent and Greiner, Markus and Keesling, Alexander and Gemelke, Nathan and Vuleti{\ifmmode\acute{c}\else\'{c}\fi}, Vladan and Kitagawa, Takuya and Wang, Sheng-Tao and Bluvstein, Dolev and Lukin, Mikhail D. and Lukin, Alexander and Zhou, Hengyun and Cant{\ifmmode\acute{u}\else\'{u}\fi}, Sergio H.},
	title = {{Experimental demonstration of logical magic state distillation}},
	journal = {Nature},
	volume = {645},
	number = {8081},
	pages = {620--625},
	year = {2025},
	month = sep,
	issn = {1476-4687},
	publisher = {Nature Publishing Group},
	url = {https://doi.org/10.1038/s41586-025-09367-3},
	doi = {10.1038/s41586-025-09367-3}
}

@article{zhao2022realization,
	author = {Zhao, Youwei and Ye, Yangsen and Huang, He-Liang and Zhang, Yiming and Wu, Dachao and Guan, Huijie and Zhu, Qingling and Wei, Zuolin and He, Tan and Cao, Sirui and Chen, Fusheng and Chung, Tung-Hsun and Deng, Hui and Fan, Daojin and Gong, Ming and Guo, Cheng and Guo, Shaojun and Han, Lianchen and Li, Na and Li, Shaowei and Li, Yuan and Liang, Futian and Lin, Jin and Qian, Haoran and Rong, Hao and Su, Hong and Sun, Lihua and Wang, Shiyu and Wu, Yulin and Xu, Yu and Ying, Chong and Yu, Jiale and Zha, Chen and Zhang, Kaili and Huo, Yong-Heng and Lu, Chao-Yang and Peng, Cheng-Zhi and Zhu, Xiaobo and Pan, Jian-Wei},
	title = {{Realization of an Error-Correcting Surface Code with Superconducting Qubits}},
	journal = {Phys. Rev. Lett.},
	volume = {129},
	number = {3},
	pages = {030501},
	year = {2022},
	month = jul,
	publisher = {American Physical Society},
	url = {https://doi.org/10.1103/PhysRevLett.129.030501},
	doi = {10.1103/PhysRevLett.129.030501}
}

@article{ryan2021realization,
	author = {Ryan-Anderson, C. and Bohnet, J. G. and Lee, K. and Gresh, D. and Hankin, A. and Gaebler, J. P. and Francois, D. and Chernoguzov, A. and Lucchetti, D. and Brown, N. C. and Gatterman, T. M. and Halit, S. K. and Gilmore, K. and Gerber, J. A. and Neyenhuis, B. and Hayes, D. and Stutz, R. P.},
	title = {{Realization of Real-Time Fault-Tolerant Quantum Error Correction}},
	journal = {Phys. Rev. X},
	volume = {11},
	number = {4},
	pages = {041058},
	year = {2021},
	month = dec,
	publisher = {American Physical Society},
	url = {https://doi.org/10.1103/PhysRevX.11.041058},
	doi = {10.1103/PhysRevX.11.041058}
}

@article{bravyi2005universal,
	author = {Bravyi, Sergey and Kitaev, Alexei},
	title = {{Universal quantum computation with ideal Clifford gates and noisy ancillas}},
	journal = {Phys. Rev. A},
	volume = {71},
	number = {2},
	pages = {022316},
	year = {2005},
	month = feb,
	publisher = {American Physical Society},
	url = {https://doi.org/10.1103/PhysRevA.71.022316},
	doi = {10.1103/PhysRevA.71.022316}
}

@article{gottesman1999demonstrating,
	author = {Gottesman, Daniel and Chuang, Isaac L.},
	title = {{Demonstrating the viability of universal quantum computation using teleportation and single-qubit operations}},
	journal = {Nature},
	volume = {402},
	number = {6760},
	pages = {390--393},
	year = {1999},
	month = nov,
	issn = {1476-4687},
	publisher = {Nature Publishing Group},
	url = {https://doi.org/10.1038/46503},
	doi = {10.1038/46503}
}

@article{gottesman2014fault,
	author = {Gottesman, Daniel},
	title = {{Fault-tolerant quantum computation with constant overhead}},
	journal = {Quantum Inf. Comput.},
	volume = {14},
	number = {15-16},
	pages = {1338--1372},
	year = {2014},
	month = nov,
	issn = {1533-7146},
	publisher = {Rinton Press, Incorporated},
	url = {https://doi.org/10.5555/2685179.2685184},
	doi = {10.5555/2685179.2685184}
}

@article{tillich2013quantum,
	author = {Tillich, Jean-Pierre and Z{\ifmmode\acute{e}\else\'{e}\fi}mor, Gilles},
	title = {{Quantum LDPC Codes With Positive Rate and Minimum Distance Proportional to the Square Root of the Blocklength}},
	journal = {IEEE Trans. Inf. Theory},
	volume = {60},
	number = {2},
	pages = {1193--1202},
	year = {2013},
	month = nov,
	publisher = {IEEE},
	url = {https://doi.org/10.1109/TIT.2013.2292061},
	doi = {10.1109/TIT.2013.2292061}
}

@article{xu2024constant,
	author = {Xu, Qian and Bonilla Ataides, J. Pablo and Pattison, Christopher A. and Raveendran, Nithin and Bluvstein, Dolev and Wurtz, Jonathan and Vasi{\ifmmode\acute{c}\else\'{c}\fi}, Bane and Lukin, Mikhail D. and Jiang, Liang and Zhou, Hengyun},
	title = {{Constant-overhead fault-tolerant quantum computation with reconfigurable atom arrays}},
	journal = {Nat. Phys.},
	volume = {20},
	number = {7},
	pages = {1084--1090},
	year = {2024},
	month = jul,
	issn = {1745-2481},
	publisher = {Nature Publishing Group},
	url = {https://doi.org/10.1038/s41567-024-02479-z},
	doi = {10.1038/s41567-024-02479-z}
}

@article{egan2021fault,
	author = {Egan, Laird and Debroy, Dripto M. and Noel, Crystal and Risinger, Andrew and Zhu, Daiwei and Biswas, Debopriyo and Newman, Michael and Li, Muyuan and Brown, Kenneth R. and Cetina, Marko and Monroe, Christopher},
	title = {{Fault-tolerant control of an error-corrected qubit}},
	journal = {Nature},
	volume = {598},
	number = {7880},
	pages = {281--286},
	year = {2021},
	month = oct,
	issn = {1476-4687},
	publisher = {Nature Publishing Group},
	url = {https://doi.org/10.1038/s41586-021-03928-y},
	doi = {10.1038/s41586-021-03928-y}
}

@article{chamberland2016thresholds,
	author = {Chamberland, Christopher and Jochym-O{'}Connor, Tomas and Laflamme, Raymond},
	doi = {10.1103/PhysRevLett.117.010501},
	journal = {Phys. Rev. Lett.},
	month = jun,
	number = {1},
	pages = {010501},
	publisher = {American Physical Society},
	title = {{Thresholds for Universal Concatenated Quantum Codes}},
	url = {https://doi.org/10.1103/PhysRevLett.117.010501},
	volume = {117},
	year = {2016}}

@article{heussen2023strategies,
	author = {Heu{\ss}en, Sascha and Postler, Lukas and Rispler, Manuel and Pogorelov, Ivan and Marciniak, Christian D. and Monz, Thomas and Schindler, Philipp and M{\ifmmode\ddot{u}\else\"{u}\fi}ller, Markus},
	doi = {10.1103/PhysRevA.107.042422},
	issn = {2469-9934},
	journal = {Phys. Rev. A},
	month = apr,
	number = {4},
	pages = {042422},
	publisher = {American Physical Society},
	title = {{Strategies for a practical advantage of fault-tolerant circuit design in noisy trapped-ion quantum computers}},
	url = {https://doi.org/10.1103/PhysRevA.107.042422},
	volume = {107},
	year = {2023}}

@misc{beni2025tesseract,
	author = {Beni, Laleh Aghababaie and Higgott, Oscar and Shutty, Noah},
	month = mar,
	title = {{Tesseract: A search-based decoder for quantum error correction}},
	archivePrefix={arXiv},
    eprint={2503.10988},
	year = {2025}}

@article{chamberland2018flag,
	author = {Chamberland, Christopher and Beverland, Michael E.},
	title = {{Flag fault-tolerant error correction with arbitrary distance codes}},
	journal = {Quantum},
	volume = {2},
	pages = {53},
	year = {2018},
	month = feb,
	publisher = {Verein zur F{\ifmmode\ddot{o}\else\"{o}\fi}rderung des Open Access Publizierens in den Quantenwissenschaften},
	eprint = {1708.02246v3},
	url = {https://doi.org/10.22331/q-2018-02-08-53},
	doi = {10.22331/q-2018-02-08-53}
}

@article{knill1998resilient,
	author = {Knill, Emanuel and Laflamme, Raymond and Zurek, Wojciech H.},
	doi = {10.1098/rspa.1998.0166},
	journal = {Proceedings of the Royal Society of London. Series A: Mathematical, Physical and Engineering Sciences},
	number = {1969},
	pages = {365-384},
	title = {Resilient quantum computation: error models and thresholds},
	url = {https://royalsocietypublishing.org/doi/abs/10.1098/rspa.1998.0166},
	volume = {454},
	year = {1998}}

@article{aliferis2006quantum,
	author = {Aliferis, Panos and Gottesman, Daniel and Preskill, John},
title = {Quantum accuracy threshold for concatenated distance-3 codes},
year = {2006},
issue_date = {March 2006},
publisher = {Rinton Press, Incorporated},
address = {Paramus, NJ},
volume = {6},
number = {2},
issn = {1533-7146},
journal = {Quantum Info. Comput.},
month = mar,
pages = {97–165},
numpages = {69},
url={https://dl.acm.org/doi/10.5555/2011665.2011666}}

@article{jochym2014using,
	author = {Jochym-O{'}Connor, Tomas and Laflamme, Raymond},
	journal = {Phys. Rev. Lett.},
	month = jan,
	number = {1},
	pages = {010505},
	title = {{Using Concatenated Quantum Codes for Universal Fault-Tolerant Quantum Gates}},
	url = {https://doi.org/10.1103/PhysRevLett.112.010505},
	volume = {112},
	year = {2014}}

@article{yoder2016universal,
	author = {Yoder, Theodore J. and Takagi, Ryuji and Chuang, Isaac L.},
	title = {{Universal Fault-Tolerant Gates on Concatenated Stabilizer Codes}},
	journal = {Phys. Rev. X},
	volume = {6},
	number = {3},
	pages = {031039},
	year = {2016},
	month = sep,
	publisher = {American Physical Society},
	url = {https://doi.org/10.1103/PhysRevX.6.031039},
	doi = {10.1103/PhysRevX.6.031039}
}

@article{chao2018fault,
  title={Fault-tolerant quantum computation with few qubits},
  author={Chao, Rui and Reichardt, Ben W},
  journal={npj Quantum Information},
  volume={4},
  number={1},
  pages={42},
  year={2018},
  publisher={Nature Publishing Group UK London},
  doi={10.1038/s41534-018-0085-z}
}

@article{old2024lift,
	author = {Old, Josias and Rispler, Manuel and M{\ifmmode\ddot{u}\else\"{u}\fi}ller, Markus},
	title = {{Lift-connected surface codes}},
	journal = {Quantum Sci. Technol.},
	volume = {9},
	number = {4},
	pages = {045012},
	year = {2024},
	month = jul,
	issn = {2058-9565},
	publisher = {IOP Publishing},
	url = {https://doi.org/10.1088/2058-9565/ad5eb6},
	doi = {10.1088/2058-9565/ad5eb6}
}

@article{gidney2021stim,
  doi = {10.22331/q-2021-07-06-497},
  url = {https://doi.org/10.22331/q-2021-07-06-497},
  title = {Stim: A fast stabilizer circuit simulator},
  author = {Gidney, Craig},
  journal = {{Quantum}},
  year = {2021},
  issn = {2521-327X},
  publisher = {{Verein zur F{\"{o}}rderung des Open Access Publizierens in den Quantenwissenschaften}},
  volume = {5},
  pages = {497},
  month = {jul},
}

@article{ai2024quantum,
  title={Quantum error correction below the surface code threshold},
  author={Quantum AI, Google and others},
  journal={Nature},
  volume={638},
  number={8052},
  pages={920},
  year={2024},
  doi={10.1038/s41586-024-08449-y}
}

@article{postler2024demonstration,
  title={Demonstration of fault-tolerant Steane quantum error correction},
  author={Postler, Lukas and Butt, Friederike and Pogorelov, Ivan and Marciniak, Christian D and Heu{\ss}en, Sascha and Blatt, Rainer and Schindler, Philipp and Rispler, Manuel and M{\"u}ller, Markus and Monz, Thomas},
  journal={PRX Quantum},
  volume={5},
  number={3},
  pages={030326},
  year={2024},
  publisher={APS},
  doi={10.1103/PRXQuantum.5.030326}
}

@article{krinner2022realizing,
  title={Realizing repeated quantum error correction in a distance-three surface code},
  author={Krinner, Sebastian and Lacroix, Nathan and Remm, Ants and Di Paolo, Agustin and Genois, Elie and Leroux, Catherine and Hellings, Christoph and Lazar, Stefania and Swiadek, Francois and Herrmann, Johannes and others},
  journal={Nature},
  volume={605},
  number={7911},
  pages={669},
  year={2022},
  publisher={Nature Publishing Group UK London},
  doi={10.1038/s41586-022-04566-8}
}

@article{postler2022demonstration,
  title={Demonstration of fault-tolerant universal quantum gate operations},
  author={Postler, Lukas and Heu$\beta$en, Sascha and Pogorelov, Ivan and Rispler, Manuel and Feldker, Thomas and Meth, Michael and Marciniak, Christian D and Stricker, Roman and Ringbauer, Martin and Blatt, Rainer and others},
  journal={Nature},
  volume={605},
  number={7911},
  pages={675},
  year={2022},
  publisher={Nature Publishing Group UK London},
  doi={10.1038/s41586-022-04721-1}
}

@misc{ryananderson2022implementingfaulttolerantentanglinggates,
      title={Implementing fault-tolerant entangling gates on the five-qubit code and the color code}, 
      author={C. Ryan-Anderson and N. C. Brown and M. S. Allman and B. Arkin and G. Asa-Attuah and C. Baldwin and J. Berg and J. G. Bohnet and S. Braxton and N. Burdick and J. P. Campora and A. Chernoguzov and J. Esposito and B. Evans and D. Francois and J. P. Gaebler and T. M. Gatterman and J. Gerber and K. Gilmore and D. Gresh and A. Hall and A. Hankin and J. Hostetter and D. Lucchetti and K. Mayer and J. Myers and B. Neyenhuis and J. Santiago and J. Sedlacek and T. Skripka and A. Slattery and R. P. Stutz and J. Tait and R. Tobey and G. Vittorini and J. Walker and D. Hayes},
      year={2022},
      eprint={2208.01863},
      archivePrefix={arXiv},
}

@article{zhou2025low,
	author = {Zhou, Hengyun and Zhao, Chen and Cain, Madelyn and Bluvstein, Dolev and Maskara, Nishad and Duckering, Casey and Hu, Hong-Ye and Wang, Sheng-Tao and Kubica, Aleksander and Lukin, Mikhail D.},
	title = {{Low-overhead transversal fault tolerance for universal quantum computation}},
	journal = {Nature},
	volume = {646},
	number = {8084},
	pages = {303--308},
	year = {2025},
	month = oct,
	issn = {1476-4687},
	publisher = {Nature Publishing Group},
	url = {https://doi.org/10.1038/s41586-025-09543-5},
	doi = {10.1038/s41586-025-09543-5}
}

@misc{malcolm2025computingefficientlyqldpccodes,
      title={Computing efficiently in QLDPC codes}, 
      author={Alexander J. Malcolm and Andrew N. Glaudell and Patricio Fuentes and Daryus Chandra and Alexis Schotte and Colby DeLisle and Rafael Haenel and Amir Ebrahimi and Joschka Roffe and Armanda O. Quintavalle and Stefanie J. Beale and Nicholas R. Lee-Hone and Stephanie Simmons},
      year={2025},
      eprint={2502.07150},
      archivePrefix={arXiv},
}

@misc{he2025extractorsqldpcarchitecturesefficient,
      title={Extractors: QLDPC architectures for efficient Pauli-based computation}, 
      author={Zhiyang He and Alexander Cowtan and Dominic J. Williamson and Theodore J. Yoder},
      year={2025},
      eprint={2503.10390},
      archivePrefix={arXiv},
}

@misc{eberhardt2024logicaloperatorsfoldtransversalgates,
      title={Logical operators and fold-transversal gates of bivariate bicycle codes}, 
      author={Jens Niklas Eberhardt and Vincent Steffan},
      year={2024},
      eprint={2407.03973},
      archivePrefix={arXiv},
}

@misc{cross2024improvedqldpcsurgerylogical,
      title={Improved QLDPC surgery: Logical measurements and bridging codes}, 
      author={Andrew Cross and Zhiyang He and Patrick Rall and Theodore Yoder},
      year={2024},
      eprint={2407.18393},
      archivePrefix={arXiv},
}

@article{Pecorari_2025,
	author = {Pecorari, Laura and Jandura, Sven and Brennen, Gavin K. and Pupillo, Guido},
	title = {{High-rate quantum LDPC codes for long-range-connected neutral atom registers}},
	journal = {Nat. Commun.},
	volume = {16},
	number = {1111},
	pages = {1--9},
	year = {2025},
	month = jan,
	issn = {2041-1723},
	publisher = {Nature Publishing Group},
	url = {https://doi.org/10.1038/s41467-025-56255-5},
	doi = {10.1038/s41467-025-56255-5}
}

@article{Bravyi_2024,
   title={High-threshold and low-overhead fault-tolerant quantum memory},
   volume={627},
   ISSN={1476-4687},
   url={http://dx.doi.org/10.1038/s41586-024-07107-7},
   DOI={10.1038/s41586-024-07107-7},
   number={8005},
   journal={Nature},
   publisher={Springer Science and Business Media LLC},
   author={Bravyi, Sergey and Cross, Andrew W. and Gambetta, Jay M. and Maslov, Dmitri and Rall, Patrick and Yoder, Theodore J.},
   year={2024},
   month=mar, pages={778–782} }

@misc{pecos,
  author = {Ryan-Anderson, Ciaran},
  title = {PECOS: Performance estimator of codes on surfaces},
  year = {2019},
  publisher = {GitHub},
  journal = {GitHub repository},
  howpublished = {\url{https://github.com/PECOS-packages/PECOS}}
}

@article{chamberland2019fault,
  title={Fault-tolerant magic state preparation with flag qubits},
  author={Chamberland, Christopher and Cross, Andrew W},
  journal={Quantum},
  volume={3},
  pages={143},
  year={2019},
  publisher={Verein zur F{\"o}rderung des Open Access Publizierens in den Quantenwissenschaften}, 
  doi = {10.22331/q-2019-05-20-143},
  url = {https://doi.org/10.22331/q-2019-05-20-143}
}

@article{goto2016minimizing,
	author={Goto, Hayato},
	title={Minimizing resource overheads for fault-tolerant preparation of encoded states of the Steane code},
	journal={Scientific Reports},
	year={2016},
	month={Jan},
	day={27},
	volume={6},
	number={1},
	pages={19578},
	doi={10.1038/srep19578},
	url = {https://doi.org/10.1038/srep19578}
}

@misc{meier2012magic,
	title={Magic-state distillation with the four-qubit code}, 
	author={Adam M. Meier and Bryan Eastin and Emanuel Knill},
	year={2012},
	eprint={1204.4221},
	archivePrefix={arXiv},
}

@book{nielsen2010quantum,
	place={Cambridge},
	title={Quantum computation and quantum information: 10th anniversary edition},
	DOI={10.1017/CBO9780511976667},
	url = {https://doi.org/10.1017/CBO9780511976667},
	publisher={Cambridge University Press},
	author={Nielsen, Michael A. and Chuang, Isaac L.},
	year={2010}
}

@article{evered2023high,
	author = {Evered, Simon J. and Bluvstein, Dolev and Kalinowski, Marcin and Ebadi, Sepehr and Manovitz, Tom and Zhou, Hengyun and Li, Sophie H. and Geim, Alexandra A. and Wang, Tout T. and Maskara, Nishad and Levine, Harry and Semeghini, Giulia and Greiner, Markus and Vuleti{\ifmmode\acute{c}\else\'{c}\fi}, Vladan and Lukin, Mikhail D.},
	title = {{High-fidelity parallel entangling gates on a neutral-atom quantum computer}},
	journal = {Nature},
	volume = {622},
	number = {7982},
	pages = {268--272},
	year = {2023},
	month = oct,
	issn = {1476-4687},
	publisher = {Nature Publishing Group},
	url = {https://doi.org/10.1038/s41586-023-06481-y},
	doi = {10.1038/s41586-023-06481-y}
}

@article{espinoza2021highfidelity,
  title = {High-fidelity method for a single-step $N$-bit Toffoli gate in trapped ions},
  author = {Arias Espinoza, Juan Diego and Groenland, Koen and Mazzanti, Matteo and Schoutens, Kareljan and Gerritsma, Rene},
  journal = {Phys. Rev. A},
  volume = {103},
  issue = {5},
  pages = {052437},
  numpages = {9},
  year = {2021},
  month = {May},
  publisher = {American Physical Society},
  doi = {10.1103/PhysRevA.103.052437},
  url = {https://link.aps.org/doi/10.1103/PhysRevA.103.052437}
}

@misc{sahay2025foldtransversalsurfacecodecultivation,
      title={Fold-transversal surface code cultivation}, 
      author={Kaavya Sahay and Pei-Kai Tsai and Kathleen Chang and Qile Su and Thomas B. Smith and Shraddha Singh and Shruti Puri},
      year={2025},
      eprint={2509.05212},
      archivePrefix={arXiv},
}

@misc{old2025faulttolerantstabilizermeasurementssurface,
      title={Fault-tolerant stabilizer masurements in surface codes with three-qubit gates}, 
      author={Josias Old and Stephan Tasler and Michael J. Hartmann and Markus Müller},
      year={2025},
      eprint={2506.09029},
      archivePrefix={arXiv},
}

@misc{josias_old_2025_17574057,
  author       = {Josias Old and
                  Heußen, Sascha},
  title        = {Data for Addressable FT Universal Quantum Gates
                   for LCS Codes: v1.0.0
                  },
  month        = nov,
  year         = 2025,
  publisher    = {Zenodo},
  version      = {v1.0.0},
  doi          = {10.5281/zenodo.17574057},
  url          = {https://doi.org/10.5281/zenodo.17574057},
}

\clearpage
\appendix

\section{LCS codes}\label{app:LCScodes}
We investigate the $(\ell,L) = (1,3)$-LCS codes with parameters $\llbracket n,k,d \rrbracket = \llbracket 15,3,3 \rrbracket$ with stabilizer generators
\begin{align*}
    S_X^{(0)} &= X_{0} X_{6} X_{7} X_{12} \\
    S_X^{(1)} &= X_{1} X_{7} X_{8} X_{13} \\
    S_X^{(2)} &= X_{2} X_{6} X_{8} X_{14} \\
    S_X^{(3)} &= X_{3} X_{9} X_{10} X_{12} X_{14} \\
    S_X^{(4)} &= X_{4} X_{10} X_{11} X_{12} X_{13} \\
    S_X^{(5)} &= X_{5} X_{9} X_{11} X_{13} X_{14} \\
    S_Z^{(0)} &= Z_{0} Z_{3} Z_{4} Z_{12} \\
    S_Z^{(1)} &= Z_{1} Z_{4} Z_{5} Z_{13} \\
    S_Z^{(2)} &= Z_{2} Z_{3} Z_{5} Z_{14} \\
    S_Z^{(3)} &= Z_{6} Z_{9} Z_{10} Z_{12} Z_{14} \\
    S_Z^{(4)} &= Z_{7} Z_{10} Z_{11} Z_{12} Z_{13} \\
    S_Z^{(5)} &= Z_{8} Z_{9} Z_{11} Z_{13} Z_{14}.
\end{align*}
The logical Pauli operators can be chosen as
\begin{align*}
    \lx_0 &= X_{0} X_{10} X_{12} \\
    \lx_1 &= X_{1} X_{11} X_{13} \\
    \lx_2 &= X_{2} X_{9} X_{14} \\
    \lz_0 &= Z_{0} Z_{10} Z_{12} \\
    \lz_1 &= Z_{1} Z_{11} Z_{13} \\
    \lz_2 &= Z_{2} Z_{9} Z_{14}.
\end{align*}
For more details, we refer to Ref.\,\cite{old2024lift}.

\section{Circuits} \label{app:circuits}

\subsection{Logical  \texorpdfstring{$\overline{S}$}{S} gate}
\begin{figure}
        \centering
        \includegraphics[width=\linewidth]{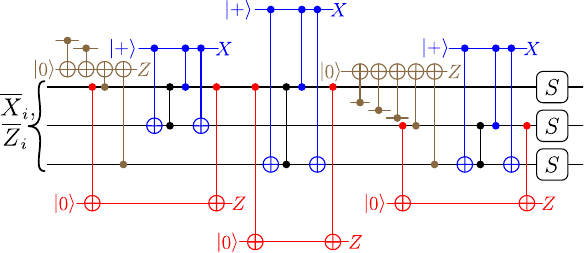}
        \caption{Circuit for the flagged, targeted logical $\ls[i]$ gate. The unflagged gate contains $ZCZ$-gates. Therefore, one uses $X$- and $Z$-flags to flag correlated faults and measures $Z$-stabilizers for input faults. }
        \label{fig:fig_flagged_s_gate}
\end{figure}

In Fig.\,\ref{fig:fig_flagged_s_gate}, we show the circuit for the flagged, targeted logical $\ls[i]$ gate. The stabilizers required to be measured are tabulated in Tab.\,\ref{tab:stab_meas}.

\subsection{Stabilizer operators measured} \label{app:stab_meas}
In Tab.\,\ref{tab:stab_meas}, we tabulate the stabilizer operators measured in each logical gate to identify the incoming faults.

\begin{table}
    \centering
    \caption{Stabilizer operators measured in gate protocols. Logical single-qubit gates require two, logical two-qubit gates four stabilizer measurements. The ordering corresponds to the ordering in the circuits given in Figs.\,\ref{fig:fig_flagging_h_gate},\ref{fig:fig_flagged_cnot_gate},\ref{fig:fig_flagged_s_gate}. The maximum weight of any stabilizer operator measured is $6$.}
    \label{tab:stab_meas}
    \begin{tabular}{lcc}
    \toprule
    gate & logical & stabilizers\\
    \midrule
                                   & $0$         & $S_Z^{(0)}$, $S_Z^{(3)}$ \\
                                   & $1$         & $S_Z^{(1)}$, $S_Z^{(4)}$ \\
    \multirow{-3}{*}{$\ls$}        & $2$         & $S_Z^{(2)}$, $S_Z^{(5)}$ \\
    \midrule
                                   & $0$         & $\prod_{i=0}^{2} S_Z^{(i)}S_X^{(i)}=Y_0Y_1Y_2Y_{12}Y_{13}Y_{14}$, $S_X^{(3)} S_Z^{(3)}$  \\
                                   & $1$         & $S_X^{(1)} S_Z^{(1)}$, $S_X^{(4)} S_Z^{(4)}$  \\
    \multirow{-3}{*}{$\lh$}        & $2$         & $\prod_{i=0}^{2} S_Z^{(i)}S_X^{(i)}=Y_0Y_1Y_2Y_{12}Y_{13}Y_{14}$, $S_X^{(5)} S_Z^{(5)}$  \\
    \midrule
                                   & $0,1$       & $S_Z^{(0)}$, $S_Z^{(3)}$, $S_X^{(1)}$, $S_X^{(5)}$ \\
                                   & $1,0$       & $S_Z^{(1)}$, $S_Z^{(5)}$, $S_X^{(0)}$, $S_X^{(3)}$  \\
                                   & $0,2$       & $S_Z^{(0)}$, $S_Z^{(3)}$, $S_X^{(2)}$, $S_X^{(5)}$  \\
                                   & $2,0$       & $S_Z^{(2)}$, $S_Z^{(3)}$, $S_X^{(0)}$, $S_X^{(4)}$  \\
                                   & $1,2$       & $S_Z^{(1)}$, $S_Z^{(5)}$, $S_X^{(2)}$, $S_X^{(3)}$  \\
    \multirow{-6}{*}{$\lcx$}       & $2,1$       & $S_Z^{(2)}$, $S_Z^{(5)}$, $S_X^{(1)}$, $S_X^{(4)}$  \\
    \bottomrule
    \end{tabular}
\end{table}

\subsection{FT magic state preparation} \label{app:magic}
In this section we discuss our adaptation of FT magic state preparation from Refs.\,\cite{goto2016minimizing, chamberland2019fault, heussen2023strategies} to the $[[15,3,3]]$ LCS code. The scheme relies on repeatedly projecting a noisy logical state onto the eigensystem of a logical Hadamard operator\,\cite{meier2012magic}. By carefully choosing our physical circuits, we can ensure that under the general depolarizing noise model of Eq.~\eqref{eq:depol} for all circuit components, no logical failure can occur in $\mathcal{O}(p)$, even though we make use of three-qubit gates in order to measure the logical Hadamard operator.

\begin{figure}
        \centering
        \includegraphics[width=\linewidth]{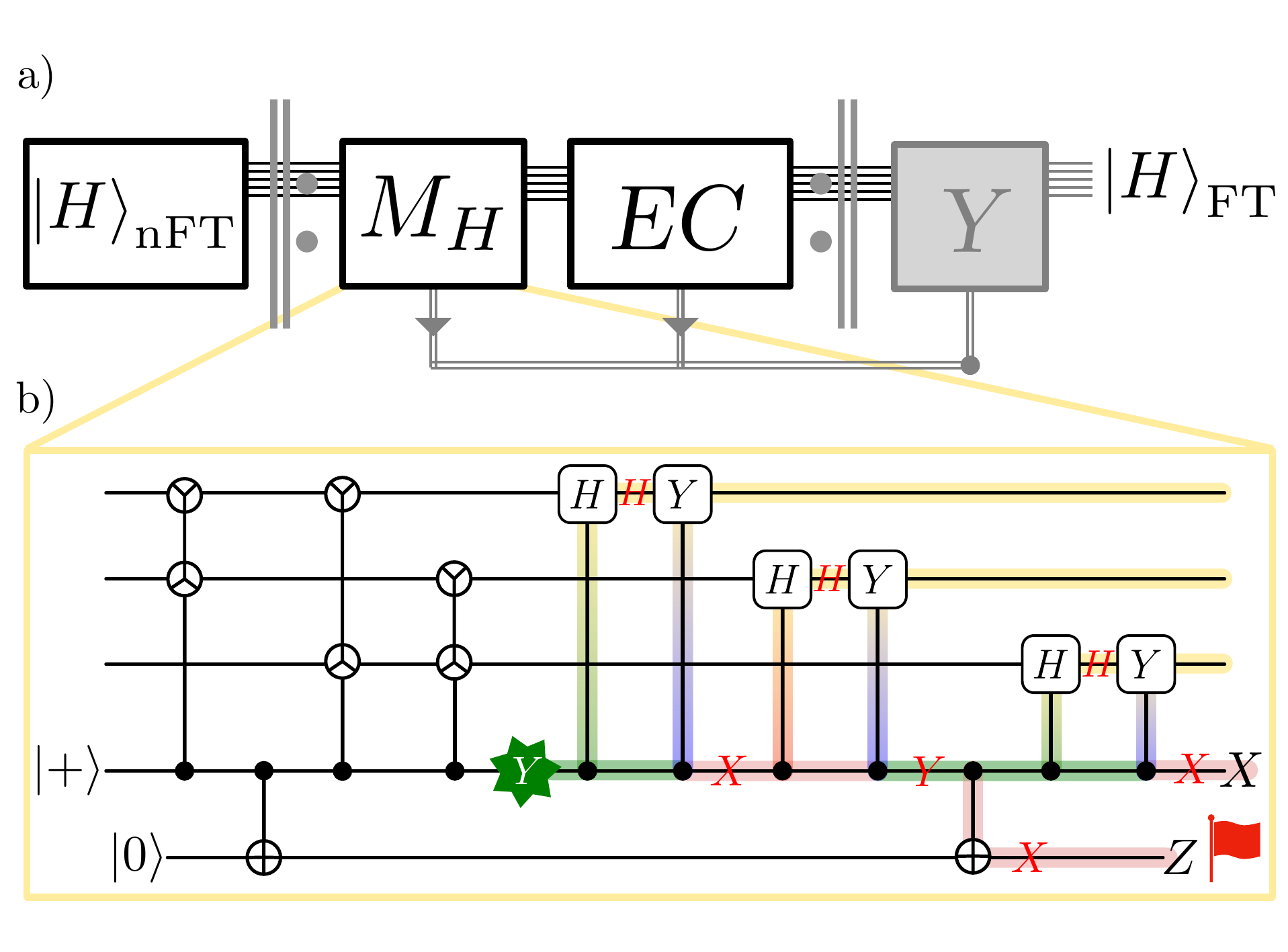}
        \caption{Fault-tolerant magic state preparation circuit. (a) Logical blocks to execute the protocol either in a deterministic way or as a repeat-until-success scheme. Only one execution of the three blocks is required per trial of repeat-until-success. To deterministically prepare a single magic state fault-tolerantly, measurement of the logical Hadamard operator $M_H$ and FT QEC need to be repeated until the measurement bit can be inferred unambiguously (see text). In case the FT measurement yields $-1$, a logical $Y$-flip is applied. (b) Circuit to measure a targeted logical Hadamard operator on three physical data qubits, on which logical operators $\overline{X}_i, \overline{Z}_i$ have their support. A single flag qubit catches otherwise undetected dangerous faults such as the one indicated by the green star. Such fault can manifest as either logical $X$ or $Z$ on the data qubits while leaving the measurement qubit unflipped.}
        \label{fig:fig_flagged_magic}
\end{figure}

The FT magic state preparation scheme is depicted on the logical level in Fig.\,\ref{fig:fig_flagged_magic}\,a). The flagged measurement circuit, shown on the physical-qubit level in Fig.\,\ref{fig:fig_flagged_magic}\,b), is always used in the first round. It is also used in the second run if previously no physical measurement was flipped neither in $M_H$ nor in the EC block. Otherwise an unflagged circuit can be used to measure the logical Hadamard operator. Only a single flag qubit is required to detect all dangerous faults, i.e., single faults that could potentially incur logical operators after performing EC. This simplification, compared to the 4-flag protocol given for the Steane code in Ref.\,\cite{chamberland2019fault}, can be attributed to the fact that, here, we only measure a weight-3 operator instead of the weight-7 logical Hadamard operator of the Steane code. 
There, several combinations of high-weight Pauli error combinations must be rendered distinguishable by a suitable flag construction. An example of a dangerous fault and its propagation to all three relevant data qubits is sketched in Fig.\,\ref{fig:fig_flagged_magic}\,b). 

Note that there are other high-weight faults in the circuit that will always be detected either by the measurement qubit or by the subsequent EC block. As an example, let us consider correlated weight-2 faults on the three-qubit controlled-$YCY$ gates. There are two cases to distinguish: 
\begin{enumerate}
    \item The propagated fault after the third controlled-$YCY$ gate is still weight-2. Then, it will not be detected by the measurement qubit since it has overlap 2 with the bitwise controlled-$H$ and the bitwise controlled-$Y$ gates. A single $X$/$Z$-fault 
    before the target qubit of a controlled-$Y$ (controlled-$H$) will (not) cause a $Z$-error on the control qubit. A single $Y$-fault before the target qubit of a controlled-$H$ (controlled-$Y$) will (not) cause a $Z$-error on the control qubit. In any case, two $Z$-flips are incurred on the measurement qubit so that the measurement in the $X$-basis will not flip. However, the EC block will detect such an error of weight 2. 
    \item The propagated fault has become weight-3 after the third controlled-$YCY$ gate. We argue that this is always a mix of $X$- and $Z$-operators and not both of them can be of weight 3. Therefore the propagated error can always be detected by the subsequent EC block. To see this, consider a fault $ZIZ$ on the data qubits after the second controlled-$YCY$ gate. This fault will be mapped to $ZYZ$ (and a $Z$ on the measurement qubit) by the subsequent controlled-$YCY$ gate, which, despite being a logical $Z$, also contains a detectable $X$-component. A similar argument holds for mixed $X$- and $Z$-faults.
\end{enumerate}

Now we discuss our measurement repetition conditions. In case two consecutive executions of $M_H$ yield the same measurement results, we take said measurement result as the FT value if, additionally, either the syndrome measured in the EC block is trivial or we have performed four repetitions of $M_H$ already. To see why the latter is necessary (compared to the typical $d=3$ repetitions in FT operator measurements), consider the case where four measurements of $\overline{H}$ have resulted in the sequence $\{+1, +1, -1, -1\}$. This sequence could be caused by a weight-2 fault occurring in the second measurement round, which collapses onto a logical operator in EC and therefore flips the measurement result in the third round. Note that we assume that, for the above sequence, the third measurement collapses a state $\overline{P} \ket{\overline{H}}$ where $P$ could be any $X, Y$ or $Z$ to $\ket{-\overline{H}}$. To distinguish this case from a simple measurement fault, the fourth repetition is necessary, which, here, will unambiguously reveal that indeed the orthogonal magic state $\ket{-\overline{H}} = \overline{Y}\ket{\overline{H}}$ is on the data qubits.

\section{Additional Simulations and Simulation Details} \label{app:add_sim}

\subsection{Pseudothresholds}
\begin{figure*}
    \centering
    \includegraphics[width=0.48\linewidth]{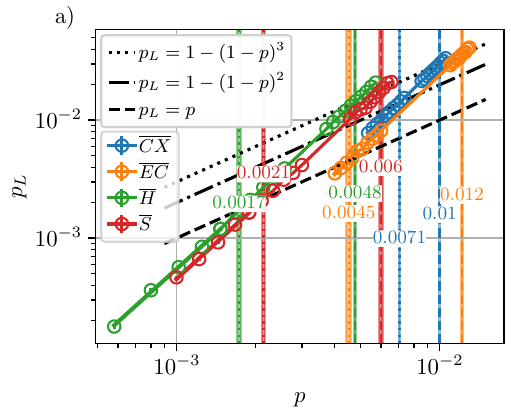}
    \includegraphics[width=0.48\linewidth]{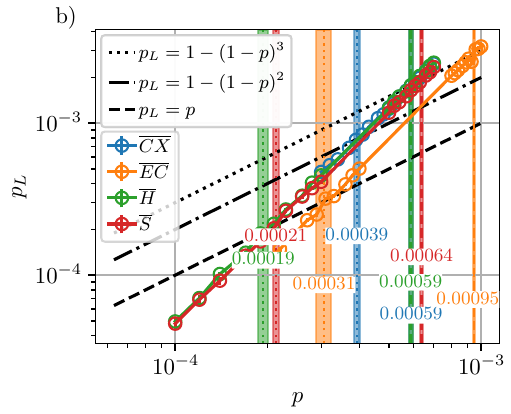}
    \caption{Details on pseudothresholds. a) In algorithmic FT, the pseudothresholds are in the order $10^{-3}$. b) In gadget FT, they are lower, $\sim 10^{-4}$.}
    \label{fig:plot_algFT_pseudo}
\end{figure*}

In Fig.\,\ref{fig:plot_algFT_pseudo} we show detailed data from which we obtain the pseudothresholds. 
We simulate around the expected crossing points and linearly interpolate the two closest points using the mean values, and a maximum and minimum estimate based on the error on the mean values. From these, we find a mean intersection, as well as an upper and a lower bound using root finding. 

\subsection{Non-averaged data}
\begin{figure*}
    \centering
    \includegraphics[width=0.3\linewidth]{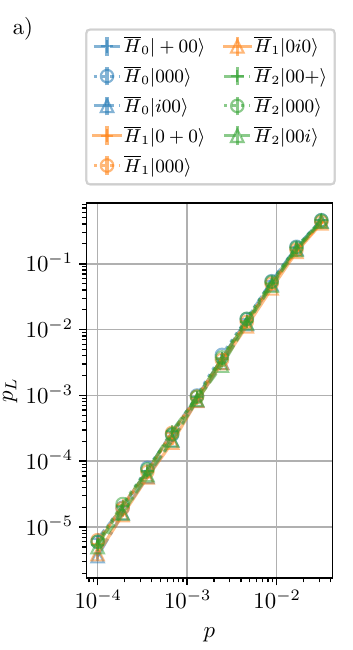}
    \includegraphics[width=0.3\linewidth]{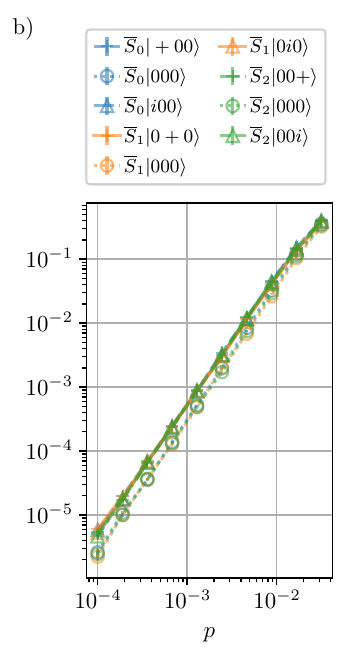}
    \includegraphics[width=0.3\linewidth]{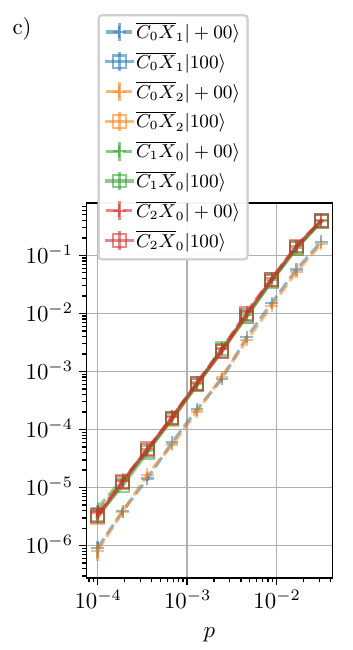}
    \caption{Flag-FT gate circuits within the algorithmic fault-tolerance framework. For the logical $\lh$ (a) and $\ls$ (b) gates, we show individual logical error probabilities for all combinations of logical operators addressed and half of the considered input states. The corresponding Pauli operator flipped input states show the same logical error rates, e.g.~$\ket{-00}$ has the same logical error rates as $\ket{+00}$. 
    As explained in the text, Pauli $Y$-eigenstates perform best for the logical $\lh$ and Pauli $Z$-eigenstates perform best for the logical $\ls$. 
    For the logical $\lcx$ (c), we only show two representative input states. Here, if the experiment has $Z$-basis input states on the control and $X$-basis on the target, the logical error rate is lower as explained in the text.}
    \label{fig:plot_algFT_nonaveraged}
\end{figure*}

In Fig.\,\ref{fig:plot_algFT_nonaveraged}, we show the raw data for the algorithmic fault-tolerance simulations before averaging. 
For all logical gates, if the input state is an eigenstate of the logical gate, it shows a lower logical error rate. This is a result of the primary logical errors induced by the circuits. 
If the all-to-all Pauli-controlled-Pauli gate (a$PCP$) part of the logical gate consists of Paulis $P$, then there is a bias towards logical $P$-errors because of the Pauli propagation. If an input state is an eigenstate of such a logical operator, it does not induce a logical error.
For the $\lh$ gate, shown in Fig.\,\ref{fig:plot_algFT_nonaveraged}\,a), these are Pauli-$Y$ input-states because $H \ket{+i/\!-\!i} \propto \ket{+i/\!-\!i}$. 
For the $\ls$ gate, shown in Fig.\,\ref{fig:plot_algFT_nonaveraged}\,b), these are Pauli-$Z$ input-states because $S \ket{0/1} \propto \ket{0/1}$. 

For the logical $\lcx$, Fig.\,\ref{fig:plot_algFT_nonaveraged}\,c), the n logical errors are biased towards of $X$-type logical errors on the target and $Z$-type logical errors on the control. 
Here, this is because an $X$-type logical error on the control propagates to the target. If the input state on the control, however, is in the $X$-basis and the input state on the target is in the $Z$-basis, these errors don't induce a logical error. This results in the lower logical error rate for numerical experiments 
$\lcx[0][1] \ket{+00}$ and $\lcx[0][2] \ket{+00}$.

\begin{figure*}
    \centering
    \includegraphics[width=0.3\linewidth]{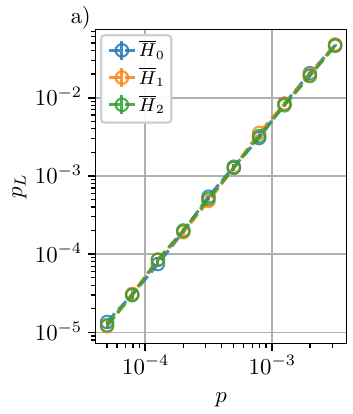}
    \includegraphics[width=0.3\linewidth]{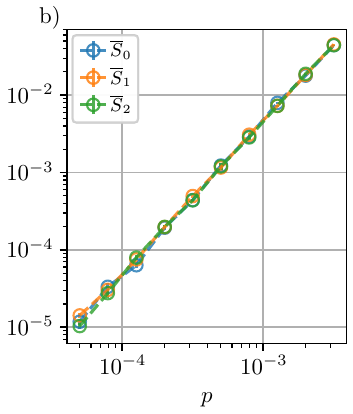}
    \includegraphics[width=0.3\linewidth]{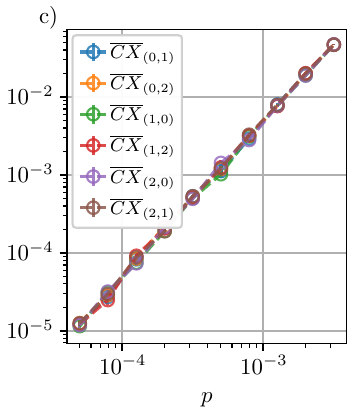}
    \caption{Flag-FT gate gadgets within the conventional fault-tolerance protocol. This protocol is independent of the input state. We show the logical error probability for all combinations of logical operators addressed. All logical gates, $\lh$ (a), $\ls$ (b) and $\lcx$ (c), have the same logical error rates within errorbars.}
    \label{fig:plot_gadgetFT_nonaveraged}
\end{figure*}

Fig.\,\ref{fig:plot_gadgetFT_nonaveraged} shows the non-averaged data for the gadget fault-tolerance simulations. As they are independent of input logical states and the gates are symmetric on different (sets of) logical qubits, they show the expected same performance over all (combinations of) logical targets.

\subsection{Logical operators measured in memory experiments}\label{app:logopmem}
In Tab.\,\ref{tab:log_meas}, we show the input state dependent logical operators measured in the memory experiments.

\begin{table}
    \centering
    \caption{Logical operators measured perfectly to obtain logical failure rates of memory experiments. Depending on the input eigenstate, different logical operators are deterministic and can therefore be measured. For single qubit gates, these are the regular transformation of Paulis through conjugation, so e.g.~if a logical qubit is initialized in $\overline{\ket{+}}$, i.e.~an $\lx$ eigenstate, the measurement after the $\lh$ gate is $\lz = $\lh$ \lx \lh$. For the logical two-qubit gate, if the control commutes with the input on the control, e.g.~\lz for a logical \lcx, the logical measurement corresponds to the input eigenstate. For anti-commutation, e.g.~\lx or \lz for the logical \lcx, we measure the joint $\overline{XX}$ and $\overline{ZZ}$ operators on the control and the target logical qubits.}
    \label{tab:log_meas}
    \begin{tabular}{lcccc}
    \toprule
                               & \multicolumn{2}{c}{input eigenstate} & \multicolumn{2}{c}{logical measurement} \\
    \multirow{-2}{*}{gate}     & control & target                     &  control & target \\
    \midrule
                                   & & $\lx$         & & $\ly$  \\
                                   & & $\ly$         & & $\lx$  \\
    \multirow{-3}{*}{$\ls$}        & & $\lz$         & & $\lz$  \\
    \midrule
                                   & & $\lx$         & & $\lz$  \\
                                   & & $\ly$         & & $\ly$  \\
    \multirow{-3}{*}{$\lh$}        & & $\lz$         & & $\lx$  \\
    \midrule
                                   & $\mid$ & $\lx$         & & $\lx$  \\
                                   & $\lz$  & $\ly$         & & $\ly$  \\
                                   & $\mid$ & $\lz$         & \multirow{-3}{*}{$\lz$} & $\lz$  \\
                                   &  $\mid$         & $\lx$         & \multicolumn{2}{c}{$\mid$}   \\
                                   &  $\lx$,$\ly$    & $\ly$         & \multicolumn{2}{c}{$\overline{XX}$,$\overline{ZZ}$} \\
    \multirow{-6}{*}{$\lcx$}       &  $\mid$         & $\lz$         & \multicolumn{2}{c}{$\mid$}  \\
    \bottomrule
    \end{tabular}
\end{table}

\section{General Logical Operator Circuits} \label{app:general_logical_ops}
\begin{figure*}
    \centering
    \includegraphics[width=\linewidth]{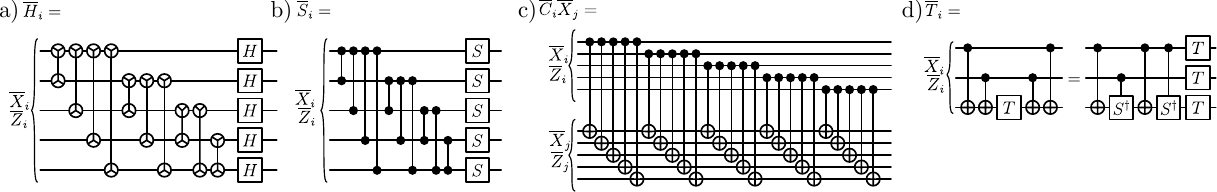}
    \caption{a) -- c) Non-FT circuits for logical gates in LCS codes of distance $d=5$. For the logical $\lh$ gate (a),  the transversal part consists only of Hadamard gates. Distances $d=3,7,11,\dots$ require another layer of transversal $Y$-gates to fix a phase, as described in the main text for general all-to-all Pauli-controlled-Pauli gates. d) A logical $T$-gate, here for $d=3$. The same decode-reencode approach can be used, and commuting the $T$-gate past the $CX$ gates on the right results in a circuit with controlled-$S^{\dagger}$ gates. }
    \label{fig:general_d_logical_gates}
\end{figure*}

In this section, we provide the general, non-FT construction of circuits for logical gates LCS codes with $d>3$. We show the circuits exemplarily for $d=5$ in Fig.\,\ref{fig:general_d_logical_gates}\,a)-c).
In general, for a code of distance $d$ and logical $\lx$ and $\lz$ operators supported on qubits $\mathcal{L}$, the single-qubit logical gate circuits are given by
\begin{align}   
    \lh &= Y_{\mathcal{L}}^{\frac{d-1}{2} \bmod 2} H_{\mathcal{L}} YCY_{\mathcal{L}}, \\
    \ls &= S_{\mathcal{L}} ZCZ_{\mathcal{L}}.
\end{align}
Here, we use the notation from the main text, i.e.~a subscript $\mathcal{L}$ on a single-qubit gates refers to its transversal implementation on the set of qubits $\mathcal{L}$. For the two-qubit gates, this is the all-to-all Pauli-controlled-Pauli on qubits $\mathcal{L}$. 
A two-qubit logical $\lcx$ from qubits supported on $\mathcal{L}_i$ and $\mathcal{L}_j$ is given by
\begin{align}
    \lcx[i][j] &= \prod_{\substack{c \in \mathcal{L}_i,\\ t \in \mathcal{L}_j \\}} C_c X_t.
\end{align}

In Fig.\,\ref{fig:general_d_logical_gates}\,d), we show a logical $\overline{T}$ gate on a distance $d=3$ code. We used the same decode-reencode approach outlined in the main text. Commuting the $T$-gate through results in a circuit with controlled-$S^{\dagger}$ gates. It remains an open question whether/how a flag construction can render this circuit fault tolerant.

\section{Scaling up round-robin flag-FT gates}\label{sec:scaleflag}

From our discussion in Sec.~\ref{sec:flaggatediscussion} and particularly our analysis of Eq.~\eqref{eq:thresha}, it is clear that our round-robin flag-FT Clifford gate constructions will retain a finite pseudothreshold when scaling up to larger distances as long as we can assure that the leading-order coefficient of logical failure rates $a$ scales at most exponentially with the code distance, 
\begin{align}
     &\mathrm{Eq.}(\ref{eq:thresha}) \implies \nonumber\\ 
     p_{\mathrm{th}} &= e^{2 \frac{\mathrm{ln}(d)}{d+1}} e^{-2 \frac{\mathrm{ln}(a(d))}{d+1}} \\&\xrightarrow{d \to \infty}\begin{cases}
         \mathrm{const.} \qif a \sim d^{\beta} e^{\alpha d} \\
         \sim e^{-d} \to 0 \qif a \sim d^{\beta} e^{\alpha d^\gamma}, \gamma > 1.
     \end{cases}
\end{align}

We now give a (rather pessimistic) estimation on the functional behavior of $a$. Let us assume that the number of dangerous fault locations for an unflagged circuit scales like the number of physical gates, which is the worst case. For a round-robin construction, the number of gates scales like $d(d+1)/2 \sim d^2$. It is reasonable to expect the number of stabilizers that have to be measured as part of the FT gate to scale at most linearly in $d$ because, in the worst case, we need to find one distinct stabilizer for each of the $d$ physical qubits in the support of a logical operator. The same applies to the number of flag qubits required for such stabilizer measurements for arbitrary distance, as proposed in Ref.\,\cite{chamberland2018flag}. Note that each stabilizer's weight is bounded by $6$ and independent of $d$ since LCS codes are LDPC; this only contributes to the prefactor but not to the scaling behavior with distance.

The arbitrary-distance flag-FT protocol of Ref.\,\cite{chamberland2018flag} requires $\mathcal{O}(d^2)$ repetitions of syndrome measurements. With the number of fault locations scaling at most like $d^2$, there are at most $\mathcal{O}\left(\binom{d^2}{t+1}\right)$ total uncorrectable fault paths. Taking a Stirling-type approximation of this binomial coefficient, we conclude that the coefficient $a$ of the leading order contribution to the logical failure rate may scale as badly as $\mathcal{O}\left((d e)^{d}\right)$. From this estimation, we find a super-exponential scaling of $a(d)$ in the limit of large distances. This may indeed prevent a finite pseudothreshold in our scheme. The number of uncorrectable faults is at most allowed to scale linearly with $d$, instead of quadratic, for $p_\mathrm{th}$ to remain finite.

While this estimate may seem discouraging, we wish to point out two aspects of putting LCS codes to use. As this code family is not asymptotically good for large numbers of qubits and high distances anyway, we should not be deterred by this result on scaling up flagged gates. The regime of interest for practical implementations currently lies in the small- to medium-distance range. Recent results, such as Ref.\,\cite{ai2024quantum}, suggest that a scale-up to very large distances may not be required after all. Furthermore, we stress that our initial assumption that \emph{all} fault locations lead to logical failure is overly pessimistic, as previous works, such as Ref.\,\cite{chamberland2018flag}, indicate that flag constructions typically act to only make a small number of problematic faults distinguishable.

\end{document}